\documentclass[titlepage,pdflatex,sn-basic]{sn-jnl}%

\usepackage{graphicx}%
\usepackage{multirow}%
\usepackage{amsmath,amssymb,amsfonts}%
\usepackage{amsthm}%
\usepackage{mathrsfs}%
\usepackage[title]{appendix}%
\usepackage{xcolor}%
\usepackage{textcomp}%
\usepackage{manyfoot}%
\usepackage{booktabs}%
\usepackage{algorithm}%
\usepackage{algorithmicx}%
\usepackage{algpseudocode}%
\usepackage{listings}%
\usepackage{enumitem}

\usepackage{hvfloat}
\usepackage{rotating}
\usepackage{todonotes} %
\usepackage{hyperref}
\usepackage{breakurl}
\PassOptionsToPackage{hyphens}{url}
\usepackage{cleveref}

\usepackage{float}
\usepackage{changepage}
\usepackage{pgfplots}
\usepackage{pgfplotstable}
\pgfplotsset{compat=1.18}
\usepackage{tikz}
\usepackage{tabularx}
\newcolumntype{C}{>{\centering\arraybackslash}X} %

\usepackage{caption}
\setcitestyle{aysep={}} 

\newcounter{definition}
\newtheorem{Definition}[definition]{Definition}

\def\mi#1{\mathit{#1}}

\makeatletter
\g@addto@macro{\UrlBreaks}{\UrlOrds}
\makeatother

\begin{document}

\title[SOPA: A Framework for Sustainability-Oriented Process Analysis and Re-design in Business Process Management]{SOPA: A Framework for Sustainability-Oriented Process Analysis and Re-design in Business Process Management}

\author*[1,2,5]{\fnm{Finn} \sur{Klessascheck}}\email{finn.klessascheck@tum.de}

\author[3,4]{\fnm{Ingo} \sur{Weber}}\email{ingo.weber@tum.de}

\author[1,2]{\fnm{Luise} \sur{Pufahl}}\email{luise.pufahl@tum.de}

\affil*[1]{\orgdiv{School of CIT}, \orgname{Technical University of Munich}, \orgaddress{\city{Heilbronn}, \country{Germany}}}

\affil[2]{\orgname{Weizenbaum Institute}, \orgaddress{\city{Berlin}, \country{Germany}}}

\affil[3]{\orgdiv{School of CIT}, \orgname{Technical University of Munich}, \orgaddress{\city{Garching}, \country{Germany}}}

\affil[4]{\orgname{Fraunhofer Gesellschaft}, \orgaddress{\city{Munich}, \country{Germany}}}

\affil[5]{\orgname{Technical University Berlin}, \orgaddress{\city{Berlin}, \country{Germany}}}

\abstract{Given the continuous global degradation of the Earth's ecosystem due to unsustainable human activity, it is increasingly important for enterprises to evaluate the effects they have on the environment. Consequently, assessing the impact of business processes on sustainability is becoming an important consideration in the discipline of Business Process Management (BPM).
However, existing practical approaches that aim at a sustainability-oriented analysis of business processes provide only a limited perspective on the environmental impact caused. Further, they provide no clear and practically applicable mechanism for sustainability-driven process analysis and re-design. Following a design science methodology, we here propose and study \textit{SOPA}, a framework for sustainability-oriented process analysis and re-design.
SOPA extends the BPM life cycle by use of Life Cycle Assessment (LCA) for sustainability analysis in combination with Activity-based Costing (ABC).
We evaluate SOPA and its usefulnesswith a case study, by means of an implementation to support the approach, thereby also illustrating the practical applicability of this work.}


\pagenumbering{gobble}

\maketitle

\vspace{-3em}

\footnotesize

\paragraph{Statements and Declarations}

\subparagraph{Competing Interests and Funding}
The authors have no relevant conflict of interest to disclose. Open access funding was provided by the Technical University of Munich. No further funds, grants, or other support was received.

\vspace{-0.3em}

\subparagraph{Acknowledgements} %
The authors thank Stephan Haarmann for his advice, which helped shape an earlier version of this work. %
Further, the authors acknowledge the contribution of Niklas Sumalvico towards the evaluation of SOPA.

\vspace{-0.3em}

\subparagraph{Ethics approval}
Not applicable.

\vspace{-0.3em}

\subparagraph{Consent to participate}
Not applicable.

\vspace{-0.3em}

\subparagraph{Consent for publication}
Not applicable.

\vspace{-0.3em}

\subparagraph{Availability of data and materials}
The datasets generated during and/or analyzed during the present study are available in a FigShare repository at \url{https://doi.org/10.6084/M9.FIGSHARE.22591513}.

\vspace{-0.3em}

\subparagraph{Code availability}
We provide the software components developed for evaluating SOPA as open-source software, available online at \url{https://github.com/INSM-TUM/sustainability-analysis-tool} and \url{https://github.com/INSM-TUM/sustainability-scylla-extension}.

\vspace{-0.3em}

\subparagraph{Authors' contributions}
Conceptualization, F.K. and L.P.; methodology, F.K., I.W. and L.P.; software, F.K.; validation, I.W. and L.P.; formal analysis, F.K.; investigation, F.K.; resources, F.K.; data curation, F.K.; writing---original draft preparation, F.K.; writing---review and editing, I.W. and L.P.; visualization, F.K.; supervision, L.P.; project administration, L.P.

\normalsize

\clearpage

\pagenumbering{arabic}

\section{Introduction}

\label{sec:intro}

The concept of sustainability has gained notable importance not only in general societal debates, but also regarding the activities of businesses.
When considering the impact of business and their operation and processes, traditionally, the focus is put on performance in a financial or operational sense~\citep{hernandezgonzalezApproachingGreenBPM2019}. 
However, the scientific consensus about the relation between human activity and the existence and acceleration of human-made climate change and alterations of ecosystems underlines the necessity of %
considering environmental impact.
The UN's \textit{Sustainable Development Goals}, released as a ``blueprint to achieve a better and more sustainable future for all''~\citep{unenvironmentGlobalEnvironmentOutlook2019}, state that
a decoupling of environmental degradation and resource usage from economic growth and associated production and consumption patterns is required.
According to the report, unsustainable human activities have significantly degraded the Earth's ecosystems on a global scale, thereby endangering the ecological foundations of society. %
Toxic substances introduced into the environment as a result of e.g.\ wasteful manufacturing, play a significant role in the endangerment of biodiversity and promotion of risks to human health --- a loss of biodiversity is closely linked to threats towards human life and its surrounding environment~\citep{romanelliConnectingGlobalPriorities2015}. Further, industrial activity and production of heat and electricity have been identified as key contributors to climate change~\citep{intergovernmentalpanelonclimatechangeClimateChange20142015}. %
Moreover, ``current goals for conserving and sustainably using nature, and achieving global sustainability, cannot be met by current trajectories'', requiring ``transformative changes across economic, social, political, and technological factors''~\citep{diazGlobalAssessmentReport2019}.

These facts underline the necessity of considering the impact of businesses and their activities on the environment, and on sustainability --- in other words, the degree to which they contribute to the ongoing deterioration of the environment and ecosystems. Only when their impact is known, business can take useful and directed measures in order to reduce that impact~\citep{hernandezgonzalezApproachingGreenBPM2019}.

In literature and in practice, various approaches have been discussed and established to assess the impact of businesses on both very broad and general scales, and with specific perspectives and narrow measures. Broad approaches include mechanisms that assess social, economic, and environmental impact of enterprises in conjunction (see e.g.\ \cite{elkingtonACCOUNTINGTRIPLEBOTTOM1998}). In the discipline of \textit{Business Process Management} (BPM), used in enterprises for modeling, analyzing, measuring and improving their business processes~\citep{weskeBusinessProcessManagement2012}, mechanisms have emerged which assess sustainability with a very limited scope on a process-level view. A well-known example is the work of \cite{reckerMeasuringCarbonFootprint2011}, where mainly carbon emissions are assessed.
While broad analyses are unsuitable for assessing and re-designing individual business processes in order to reduce their environmental impact, existing process-focused approaches generally concentrate on a limited set of measures to assess the environmental impact (e.g., carbon emissions and energy consumption~\citep{fritschPathwaysGreenerPastures2022}). This, however, discounts many dimensions along which the environment is also impacted negatively, such as emission of other greenhouse gasses that contribute to climate change, land use, aquatic acidification, or other substances toxic to humans~\citep{klopfferLifeCycleAssessment2014}. 
Nonetheless, existing literature underlines that integrating sustainability into BPM is a worthwhile research effort in terms of research opportunity and relevance~\citep{stolzeSustainabilityBusinessProcess2012}. 
For instance, enabling analysts and managers to consider sustainability on a process level leads to meaningful insights for process re-design, as the process perspective relates to where exactly impacts are caused and reductions of impact can be implemented~\citep{reckerModelingAnalyzingCarbon2012}. Further, it has been shown that enterprises have a need for techniques that enable modeling, monitoring, and optimizing the environmental impact of business processes, in order to establish sustainable business practices~\citep{sohnsGreenBusinessProcess2023}. 
However, a mechanism that operates on a process level while taking into account a holistic view on environmental sustainability, and that provides mechanisms for sustainability-oriented process analysis and re-design, does not yet exist.%

\medskip
\noindent
In light of this, we pose three research questions to be addressed by this article that specifically target this gap:
\medskip
\begin{enumerate}[label=RQ\arabic*:,wide=0pt, leftmargin=*]
	\item \label{RQ1} What is a mechanism with which the impact on environmental sustainability of business processes can be analyzed holistically, without limiting the dimensions of impact that are being considered to a small subset of indicators?
    \item \label{RQ2} How can a software system be derived from this mechanism?
	\item \label{RQ3} How can this mechanism be leveraged in a practical setting to facilitate both process analysis and re-design based on a software system?
\end{enumerate}

\bigskip
\noindent
This work targets these research questions by proposing \textit{SOPA}, a framework for sustainability-oriented process analysis, that combines \textit{Life Cycle Assessment} (LCA) and \textit{activity-based costing} (ABC) within the BPM life-cycle.
LCA is an assessment technique from sustainability research, specifically the field of \textit{environmental impact assessment} (EIA), which is concerned with systematically ``identifying, predicting, evaluating and mitigating biophysical, social and other relevant effects of proposed development proposals''~\citep{glassonIntroductionEnvironmentalImpact2019}. LCA focuses explicitly on identifying the cradle-to-grave (meaning from materials procurement until the final disposal, i.e., across the complete life cycle) environmental impact of products, services, and systems. This is in contrast to considering solely a particular location or facility, as is the case with traditional EIA~\citep{glassonIntroductionEnvironmentalImpact2019}. Notably, the integration of LCA into BPM has been identified as a key research gap~\citep{fritschPathwaysGreenerPastures2022}.
Activity-based costing is a model of determining (usually financial) costs in an organization by determining the involved activities, their respective costs, and thus the cost of the actual product~\citep{cooperDesignCostManagement1991}. This allows a more precise estimation of process costs than previous accounting methods~\citep{leeApplicationSimulationTechnique2001}. 
In SOPA, we use this notion to assess the environmental impact of individual activities, and hence, the overall process.
Thus, SOPA allows process stakeholders to assess the environmental impact of a business process holistically, and re-design it accordingly.

Concretely, SOPA contributes methods for assessing the impact of business processes on environmental sustainability by assigning impact measures based on LCA to activities with ABC, and proposes concrete measures for sustainability-oriented process re-design via business process simulation.
In contrast to existing approaches, SOPA aims at a holistic consideration of the environmental impact through LCA, instead of focusing on a few explicit impact dimensions, and more explicitly covers mechanisms for re-design.
In this work, we follow a design science research methodology~\citep{wieringaDesignScienceMethodology2014,peffersDesignScienceResearch2007} in the design and evaluation of the artifact, SOPA.
The evaluation is based on business process simulation~\citep{pufahlDesignExtensibleBPMN2018} and a prototypical implementation, and is applied to a real-world case study. We show how SOPA is able to be leveraged for process re-design by evaluating various re-design scenarios for their environmental impact and determining the least impactful variant.

The following section presents background information on sustainability, ABC, LCA, BPM and business process simulation. Then, the methodology of design science, used in this work, is detailed (\autoref{sec:methodology}). Subsequently, the resulting framework of SOPA is presented and evaluated (\autoref{sec:results}). Finally, the results, as well as practical and theoretical considerations for BPM, are discussed (\autoref{sec:discussion}).

\section{Background and Related Work}
\label{sec:background}

\subsection{Sustainability}

The term of sustainability has been established in discussions about viable long-term development, be it economic, social, or ecological.\ In 1987, the \textit{Brundtland Report} of the World Commission on Environment and Development from the United Nations formulated sustainability as ensuring that the needs of the present are met without compromising the ability of future generations to meet their own needs~\citep{brundtlandOurCommonFuture1987}.
Since then, some estimated three-hundred definitions of sustainability and sustainable development have emerged in environmental management and other disciplines~\citep{johnstonReclaimingDefinitionSustainability2007}. This underlines that, while being a term of everyday use, defining the concept itself is complex, and a general understanding and uniform definition is still missing~\citep{purvisThreePillarsSustainability2019}. Some authors, such as \cite{johnstonReclaimingDefinitionSustainability2007} argue, therefore, that an unambiguous definition of sustainability and an authoritative framework that can be used to assess policies and practice w.r.t. sustainability are needed.%

\begin{figure}[H]
  \centering
  \includegraphics[width=0.3\textwidth]{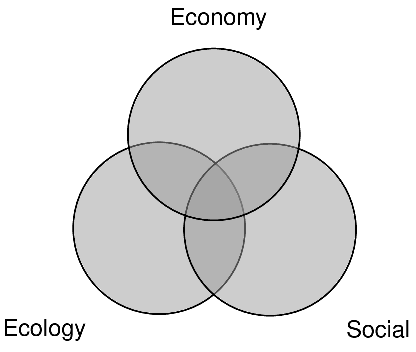}
  \caption[Triad model of sustainability]{Triad model of sustainability, adapted from \cite{giddingsEnvironmentEconomySociety2002}}
  \label{fig:sustainability}
\end{figure}

Nonetheless, a common understanding has emerged in literature, which can be summarized as the \textit{triad model of sustainability}. This model, displayed in Fig.~\ref{fig:sustainability}, illustrates that sustainability can be understood to consist of three interconnected facets, the \textit{economic sustainability}, the \textit{social sustainability}, and the \textit{environmental sustainability}. The triad nature of the model underlines that all three facets are inextricably linked, implying a strong reciprocal influence~\citep{bauerLeitbildNachhaltigenEntwicklung2008}.%

Following \cite{bauerLeitbildNachhaltigenEntwicklung2008}, the three facets have been defined as follows:
The \textit{economic} facet of sustainability is concerned with controlled growth and long-term gains in a sense of capital preservation through careful use of natural resources. However, quantitative growth is still seen as a requirement. 
The focus of \textit{social} sustainability lies on intergenerational and intragenerational justice in terms of access to resources and satisfaction of basic needs. Besides that, it explicitly touches on issues of gender relations in the sense of creating fairer living environments for all genders.
\textit{Environmental} sustainability deals with the preservation of the natural basis of life and its life cycles, i.e., ecosystems and other natural functions, through the prudent use of natural resources. This not only includes the economic necessity of preserving existing natural capital, but also securing the ecological conditions of human survival in general.

It has been argued that the difference between considering all facets at once, and only considering one facet with a limited set of indicators is a difference in perspective, of a ``\textit{broad}'' and ``\textit{narrow}'' view on sustainability~\citep{huetingBroadSustainabilityContra2004}. This work adapts a narrow perspective on sustainability, focusing on the environmental facet being impacted by business process. In other words, the main focus of SOPA lies in assessing the cost incurred by individual business processes w.r.t. the environment.

\subsection{Life Cycle Assessment}
\label{sec:background-lca}

The concept of sustainability implies that the costs of business activities and business processes are not just financial costs, but also costs towards the environment due to, e.g., emissions, energy usage, and waste. In order to capture these impacts, techniques such as \textit{life cycle assessment} (LCA) have emerged. LCA is a technique belonging to the field of \textit{environmental impact assessment} (EIA). EIA is concerned with the systematic identification, prediction, and evaluation and mitigation of biophysical (i.e., environmental), social and other impacts of development proposals~\citep{glassonIntroductionEnvironmentalImpact2019}. In contrast to other traditional EIA techniques, the focus of LCA lies in identifying the environmental impact across the entire life cycle of a product, service, or system, and not only the impact at an individual facility or location~\citep{glassonIntroductionEnvironmentalImpact2019}. LCA is further described by a set of ISO norms~\citep{ISO1404020062006}. %

\begin{figure}[htb]
    \centering
    \includegraphics[width=0.4\textwidth]{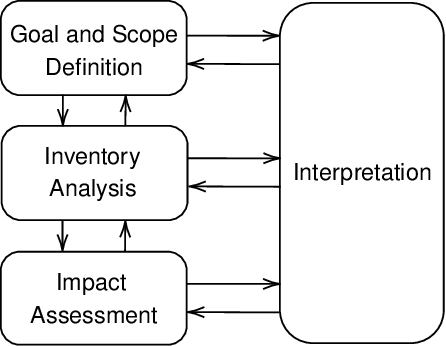}
    \caption[Life cycle assessment framework and its four phases]{Life cycle assessment framework and its four phases, adapted from~\cite{debruijnHandbookLifeCycle2002,ISO1404020062006}}
    \label{fig:lca}
\end{figure}

Sustainability analyses with LCA are conducted in four distinct phases, as illustrated in Fig.~\ref{fig:lca}.
During the \textit{goal and scope definition phase}, the analysis goal and its boundaries are defined. In terms of goals, LCA is suitable to answer several types of questions about systems: 
\textit{predictive} scenarios (how will the system behave), \textit{explorative} scenarios (how can the system behave), and \textit{normative} scenarios (how can a specific objective be reached)~\citep{finnvedenRecentDevelopmentsLife2009}.
In the \textit{life cycle inventory analysis phase} (LCI), relevant data is collected and a pre-determined calculation method is applied, quantifying all relevant input and output. Relevant units could be, e.g., \textit{kg} of carbon dioxide, cubic meters of natural gas, \textit{mg} of phenol, and others~\citep{debruijnHandbookLifeCycle2002}. The LCI phase is, generally, data, labor, and time-intensive. To alleviate this issue and aid analysts, LCI databases have been developed that contain data for products and services needed in many LCAs, such as electricity generation, raw materials, and transportation processes. They are public national or regional databases, as well as industry-provided or consultant databases, and are often-times offered with LCA software, such as the Swiss ecoinvent database~\citep{wernetEcoinventDatabaseVersion2016,finnvedenRecentDevelopmentsLife2009}.
The \textit{life cycle impact assessment phase} (LCIA) is where the results of the LCI are further processed and interpreted with respect to environmental impacts. A list of impact categories is defined (e.g., degree of ozone layer depletion, ecotoxicity, human toxicity, climate change) and models for relating the environmental interventions identified in the LCI phase to indicators for these categories are selected, after which actual results are calculated. The perspective of LCIA is holistic in nature, aiming to capture all possible areas of impact~\citep{finnvedenRecentDevelopmentsLife2009}. Further, the results can be normalized, and the different category indicators can be grouped and weighted to, e.g., aggregate them into a single score, expressing in a unit-less manner the overall environmental impact of the considered system. Some methodologies (e.g., \textit{Eco-indicator 99}~\citep{goedkoopEcoindicator99Methodology2007}) do this per default, for others, it can be done a posteriori~\citep{debruijnHandbookLifeCycle2002}.
During the \textit{life cycle interpretation phase}, the results of the analysis, as well as the underlying choices and assumptions made in each phase through the analyst, are evaluated. Overall conclusions are drawn, including an evaluation of the results w.r.t. their consistency, completeness, and robustness. Finally, conclusions and recommendations are formulated~\citep{debruijnHandbookLifeCycle2002}.

LCA has been identified as a useful tool to holistically measure the degree of impact of products, goods, or services on the environment, and thus, their impact on environmental sustainability. %

\subsection{Business Process Management}

For enabling an analysis and re-design of business process for their environmental impact, the use of LCA methods needs to be embedded into a rigorous methodological context. To this end, this section presents the discipline of \textit{business process management} (BPM), dealing with analysis, design, and implementation of business processes.
This discipline aims at improving and running business processes through a combination of technological and management sciences~\citep{vanderaalstProcessMining2016}.  

\textit{Business processes} are considered to be sets of activities that are performed, in a coordinated manner, in an organizational and technical environment, to achieve a business goal~\citep{weskeBusinessProcessManagement2012}. An example of such a process is the production, shipping, invoicing and payment of a specific good, e.g., a laptop computer.
BPM comprises concepts, techniques, and methods for supporting design, administration, configuration, enactment, as well as the analysis of business processes~\citep{weskeBusinessProcessManagement2012}.
These processes can be represented as \textit{business process models}, for example via modeling languages such as \textit{Business Process Model and Notation} (BPMN)~\citep{omg2011bpmn}. The models serve as a medium to communicate requirements and behaviors of the process they capture, and therefore serve as a central artifact of various disciplines in the area of process science~\citep{weskeBusinessProcessManagement2012}. %
Business process management systems can, based on such a process model, execute and enforce the respective business process in individual process instances~\citep{vanderaalstProcessMining2016,weskeBusinessProcessManagement2012}.

\begin{figure}[H]
    \centering
    \includegraphics[width=0.4\textwidth]{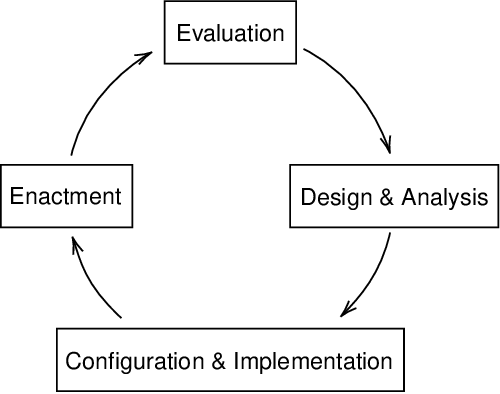}
    \caption[Business process life cycle]{Business process life cycle, adapted from \cite{weskeBusinessProcessManagement2012}}
    \label{fig:lifecycle}
\end{figure}

The business process life cycle can be structured in four main phases, as depicted in Fig.~\ref{fig:lifecycle}, with which organizations can implement and maintain business processes. 
During the \textit{design and analysis phase}, business processes are identified, analyzed, and expressed explicitly in the form of process models. Validation (i.e., ensuring that the model correctly reflects the real-world execution through, e.g., stakeholder interviews), verification (i.e., ensuring formal correctness of the model), and simulation techniques (e.g., stochastic technique to simulate process execution, as further explained in \autoref{sec:background-simulation}) can be used during this phase. The resulting process model and the behavior that it specifies are then used in the \textit{configuration and implementation phase} to implement the respective business process. This is commonly done either on a technical level by, e.g., configuring the involved information systems that support the execution of that process, by adapting business process management systems used to enact the process based on its model, or by formulating a set of rules and policies to be followed by employees.
During the \textit{enactment phase}, process instances are enacted, or executed. Here, execution data of process instances is typically gathered in the form of execution logs, or \textit{event logs} \citep{weskeBusinessProcessManagement2012}. These logs form the basis of the subsequent \textit{evaluation phase}. In this phase, the process is evaluated for shortcomings or potential for improvement, based on information captured during enactment. This phase is also related to simulation techniques of the design phase. If changes to the process need to be implemented after evaluation, a new iteration of the business process life cycle can be started, by re-entering the (re-)design phase~\citep{weskeBusinessProcessManagement2012}. Re-design objectives commonly include improvement of process performance among the dimensions of time, cost, quality, and flexibility~\citep{dumasFundamentalsBusinessProcess2018}. %

\subsubsection{Business Process Simulation}
\label{sec:background-simulation}

A method that has shown practical benefit for process analysis and re-design is \textit{business process simulation} (BPS). It can be applied to understand the behavior of a system, in order to assess the implications of alterations made to the process itself or the surroundings of the process~\citep{freitasProcessSimulationSupport2015,aguirreCombinationProcessMining2013}.
Further, it is a quantitative analysis method of process models, grounded in probability-based simulation of discrete events. BPS can be used, e.g., to consider process costs of different process alternatives~\citep{freitasProcessSimulationSupport2015}. Moreover, insights into existing or future scenarios of the execution of that process can be attained~\citep{vanderaalstBusinessProcessSimulation2010}.
BPS has been identified as an important tool in the life cycle of BPM (see Fig.~\ref{fig:lifecycle}), especially the (re-)design phase: it is possible, through simulation, to create hypothetical scenarios of process re-designs and evaluate and improve them, before implementing them in the real world~\citep{aguirreCombinationProcessMining2013}. This eliminates cost and risks inherent with testing changes in a practical setting~\citep{freitasProcessSimulationSupport2015,satyalAB-BPM-Methodology2019}.

To facilitate this simulation and analysis, a \textit{business process simulator} is required~\citep{pufahlDesignExtensibleBPMN2018}.
This tool aims at simulating the execution of process instances based on input provided by a process analyst (e.g., resource information, branch probability, number of processes instances to simulate, and other parameters~\citep{martinUseProcessMining2014}), which is usually done in the form of a \textit{simulation configuration}. The result of that simulation is artificial information in the form of, e.g., an event log and a performance report about the process, which provides, among others, information about probable throughput time, process costs and resource usage of the current process design~\citep{dumasFundamentalsBusinessProcess2018,wynnBusinessProcessSimulation2008}.
The resulting information can then be used to iteratively drive the re-design of the considered process.
However, it should be noted that BPS abstracts from the real-world behavior of the considered process, and, due to its stochastic and qualitative nature, depends greatly on the input parameters. Thus, they should be chosen with care and, if need be, validated externally~\citep{dumasFundamentalsBusinessProcess2018}.
An alternative approach is the use of historic data from logs~\citep{satyalAB-BPM-Methodology2019}, which can however only deliver reliable results if the process version from which the data stems is similar enough to the version that is simulated.
In practice, some frameworks for process re-design explicitly relying on BPS have been outlined or implemented (e.g.,~\cite{mansarBestPracticesBusiness2005}) -- however, none of them include a notion of sustainability-oriented process re-design, which is a shortcoming that SOPA aims to address.

\subsubsection{Activity-based Costing}

When considering (financial) costs involved in organizational settings, such as business processes, the perspective of \textit{activity-based costing} (ABC) has been established --- it recognizes that all activities taking place in an organization directly or indirectly support the overall business goal, and hence should be the places where costs are allocated~\citep{goebelActivityBasedCosting1998}. Thus, ABC is a model of determining (financial) costs in an organization by determining the involved activities, their respective costs, and thus determining the costs of the product or service based on the costs of the activities~\citep{cooperDesignCostManagement1991}. This allows for a more precise estimate of process costs than traditional accounting methods~\citep{leeApplicationSimulationTechnique2001}.

\begin{figure}[H]
    \centering
    \includegraphics[width=0.4\textwidth]{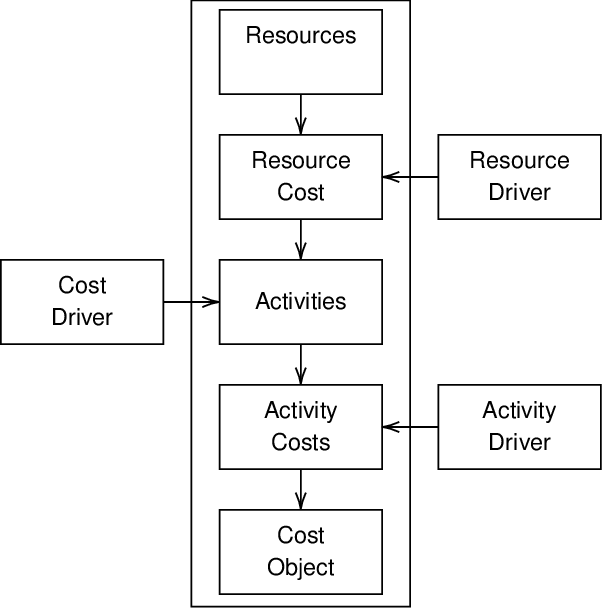}
    \caption[Structure of an activity-based costing model]{The structure of an activity-based costing model, adapted from \cite{leeApplicationSimulationTechnique2001}}
    \label{fig:abc}
\end{figure}

ABC contains several key concepts, which are illustrated in Fig.~\ref{fig:abc}.
\textit{Cost objects} represent the overall cost of the system under investigation, such as a product or a process. Activities that contribute to that product or process consume \textit{resources}, which, through \textit{resource drivers}, incur \textit{resource costs}. In the context of BPM, resources are usually understood to be, among others, human resources, machines, vehicles, or materials, which greatly contribute to the execution of individual activities and process instances~\citep{ihdeFrameworkModelingExecuting2022} --- this is a perspective similar to that in ABC~\citep{goebelActivityBasedCosting1998}.
Resource drivers represent the demand placed on resources by an activity~\citep{leeApplicationSimulationTechnique2001}, and could be, e.g., the time an employee has spent on a specific activity, depreciation of the equipment involved in that activity, the supplies used, etc. Thus, by considering what resources contribute to an activity, and how the activity places demands on them, the resource cost of that activity can be determined.
\textit{Cost drivers} determine the cost of the operational activities to which they are linked. They could be, e.g., the number of products ordered, number of parts required, etc.~\citep{leeApplicationSimulationTechnique2001}. By considering cost drivers of activities and resource costs imposed on resources by an activity, the \textit{activity costs} can be determined.
\textit{Activity drivers} represent the demand placed on activities, e.g., the number of times an activity needs to be executed, or the duration of it. With these activity drivers, the actual activity cost can be allocated to the overall cost object \citep{goebelActivityBasedCosting1998,leeApplicationSimulationTechnique2001}. Activity drivers could be, for example, the number of deliveries made to a customer, thus determining the overall cost for the customer in question~\citep{goebelActivityBasedCosting1998}.
When all activity costs have been considered and allocated to the cost object, the overall cost of the system has been determined.

In summary, ABC outlines a method in which, by considering the costs of single activities and how they relate to, e.g., a process being subject to an ABC analysis, the overall cost of that process can be determined.
SOPA uses that notion for the assessment of environmental impact: by determining the individual environmental cost of activities with LCA, the overall environmental cost of the process is determined.

\subsection{Related Work}
\label{sec:background-related}

\subsubsection{Sustainability and BPM}

The integration of sustainability considerations into the framework of BPM (also known as \textit{green BPM}) has been investigated at various levels in the literature, which is exemplified by several research streams.

\paragraph{Conceptual Integrations.}
Various contributions, particularly in \cite{vombrockeGreenBusinessProcess2012}, consider sustainability and its integration into BPM. For instance,~ \cite{seidelGreenBusinessProcess2012} provide a framework for BPM research and practice with a focus on sustainability and outline challenges for sustainable considerations along the key capabilities of BPM. For example, methods for the analysis, design, and implementation of sustainable business processes are seen as a key challenge. 
Further, \cite{brooksUnpackingGreenReview2012} investigate existing literature on IS research with a focus on sustainability, and identify the motivation for and adoption of sustainability initiatives as major research gaps.
Moreover, \cite{hovorkaInformationSystemsEnvironmental2012} critique existing sustainability analyses that focus on individual organizations, noting that they should also include other environmental factors such as biodiversity and others, and argue for challenging notions of productivity, cost reduction and profitability.
\cite{zeiseMeasurementSystemsSustainability2012} investigate performance measurement systems for their potential to be used for assessing the sustainability of companies. %
They propose a combination of reporting and management frameworks, and conclude that concrete obligatory requirements towards enterprises are needed to be able to establish concrete sustainability objectives.
While these works have contributed by integrating sustainability considerations into the BPM discipline, and have investigated subsequent capabilities, challenges, and broader concerns, they do so in a high-level manner that does not provide any actionable mechanism for analyzing and improving particular business processes. However, subsequent contributions have indeed addressed this, as we discuss in the following.

\paragraph{Structure-Based Assessments.}

The second stream, which aims to more concretely assess business processes for their environmental impact, does so based on process structures and patterns. For example, \cite{lubbeckeGuidelinesModelingEcologyAware2018,lubbeckeMethodEcologicalProcess2018} describe how business process models of certain domains (e.g., public administration) can be analyzed for pre-determined patterns that hint at a high environmental impact, and provide means for re-designing the model accordingly.
This kind of approach allows process experts to find and implement actionable ways of improving a business process. However, no quantification of impact, resp. reduced impact, is done. A further issue lies in the fact that process experts cannot readily compare and judge different process redesigns according to these patterns due to a lack of indicator-based quantification of environmental impact behind the process.

\paragraph{Indicator-Based Assessments.}
The third research stream has developed concrete indicator-based mechanisms that aim at assessing and improving the environmental impact of business processes. Accordingly, \cite{clevenManagingProcessPerformance2012} argue that techniques for measuring and managing process performance can be used to assess and improve the sustainability of enterprises.
For instance,~\cite{houyAdvancingBusinessProcess2012} propose a semi-automated approach for reducing the environmental impact of business processes. They present a method where process activities can be annotated with ratios for e.g, consumption of energy or fuel, and production of carbon dioxide with are not to be exceeded during process execution, and which can be monitored. For process re-design, they outline a mechanism that uses a library of best-practice fragments to replace ``unsustainable'' parts of a business process, to reduce e.g., carbon dioxide emissions and to not exceed the annotated ratios. 
\cite{reckerModelingAnalyzingCarbon2012} propose an extended BPMN notation to capture the consumption of fuel and paper (which both produce carbon dioxide), as well as the flow and emission of carbon dioxide in process models. Based on an ABC-like analysis of greenhouse gas emissions, the environmental impact of the process and of activities in terms of greenhouse gas emissions can be assessed, and consequently improved. \cite{wesumperumaFrameworkMultidimensionalBusiness2011,wesumperumaGreenActivityBased2013} outline a mechanism that similarly follows an ABC-based approach, focussing on greenhouse gas emissions and labor cost. Further, they propose to use simulation-based techniques to determine how activity timings change the overall greenhouse gas emissions and financial cost of the process.
Finally, the work of \cite{nowakArchitectureMethodologyFourPhased2011} describes how environmental impact in terms of greenhouse gas emissions, water usage and energy consumption can be measured or estimated during process execution on an activity level and highlighted in a process model, so that potential process alternatives can be derived. 

While this line of research and the corresponding contributions allow an assessment of business processes for their environmental impact based on specific indicators, we see a particular challenge regarding the way sustainability and environmental impact is measured.

\paragraph{Challenges of Measuring Sustainability.}
After having briefly presented extant work that integrates sustainability into BPM, especially by using environmental indicators, we now provide more detail on how sustainability is commonly measured in these works, and explicate the research gap we aim to address with SOPA.
Several structured literature reviews, such as \cite{hernandezgonzalezApproachingGreenBPM2019,roohygoharEnvironmentalSustainabilityGreen2020}, and a systematic mapping study \citep{couckuytGreenBPMBusinessOriented2019} of green BPM find that, in terms of indicators for assessing environmental impact in the BPM context (either during the design or the evaluation phase), greenhouse gas emissions, energy consumption, and consumption of specific materials such as water, are used 
(e.g., \cite{wesumperumaFrameworkMultidimensionalBusiness2011,wesumperumaGreenActivityBased2013,nowakArchitectureMethodologyFourPhased2011,houyGreenBPMSustainability2011} and \cite{reckerMeasuringCarbonFootprint2011,reckerModelingAnalyzingCarbon2012}). These impacts are usually elicited either on a process level, or on an activity basis with variations of ABC.%

In line with this, in a systematic tertiary literature review~\cite{fritschPathwaysGreenerPastures2022} identify a structured integration of LCA and BPM to incorporate sustainability considerations into BPM in a practical manner as a research gap, and share the observation of a narrow focus on carbon emissions and energy use found in related work.
While important categories of impact, they arguably do not sufficiently cover the impact that business process have on the environment.
As \cite{vanderkolkNumbersSpeakThemselves2022} argues, reducing ``environmental performance'' to a limited set of indicators such as carbon dioxide emissions and water footprint brings the danger of leading to improvement of these indicators only, ignoring the impact a process might have on environmental dimensions that are not considered by these indicators.

In summary, existing approaches have indeed contributed towards integrating aspects of sustainability into BPM, either on a conceptual level, or with techniques to consider the environmental impact of concrete business processes. However, we see that previous contributions that provide these concrete mechanisms either provide no quantification of environmental impact that would help process experts compare process variants and re-designs, or are limited to a small subset of indicators of environmental impact. This narrow focus means that related work provides only a limited picture of the environmental impact of business processes, and highlights the need for a more holistic manner of assessing the environmental impact of business processes, taking into account a broader set of impact dimensions.

Consequently, by considering related work, which we found during a literature search, and systematic literature reviews in the area of sustainability and BPM, we observe that none of the approaches provide explicit mechanisms that \emph{holistically} assess the environmental impact and provide a concrete methodology to re-design the process in a sustainability-oriented manner. 
SOPA aims to close this gap found in the literature by explicitly including the technique of BPS during the analysis and re-design phase, which is explained in the subsequent section, and by considering a more encompassing set of impact factors through the mechanism of LCA. Instead of static analysis based on just a process model, SOPA includes dynamic analysis based on process executions.

\subsubsection{Process Optimization}

In a similar vein, although not explicitly targeted towards environmental sustainability, related work in the area of BPM has investigated \textit{manual} and \textit{algorithmic} approaches for optimizing business processes towards certain goals. 
As an example for manual optimization approaches, \cite{dumasProcessRedesign2018} describe various heuristics (i.e., general ``rules of thumb'') with which improved process designs can be determined, and their impact on time, cost, quality, and flexibility of the process.
On the algorithmic side, \cite{georgoulakosEvolutionaryMultiObjectiveOptimization2017} propose a method to derive process designs that are optimal in terms of costs and durations based on activities and resources and their attributes, such as task costs and resource in- and outputs. As another example, \cite{lowRevisingHistoryCostinformed2016} present a framework to identify optimal process executions based on cost functions for cases, activities, and resources, to identify trade-offs between time, cost, and resource allocation.

While algorithmic approaches aim at automatically identifying improved process configurations that are optimal based on some parameters identified a priori, we see a challenge in applying these techniques in situations where the way in which parameters can be changed is not known beforehand and only emerges iteratively during process improvement. With SOPA we aim to allow process experts to evaluate and deliberate different improvement scenarios that are, potentially iteratively and interdependently, derived from domain knowledge. Therefore, SOPA is more closely aligned with the manual heuristics of \cite{dumasProcessRedesign2018}. For SOPA, a reduction in environmental impact through changes of activities and process implementation is the main objective, although based on a quantitative evaluation of the environmental impact and its reduction.

\section{Methods}

\label{sec:methodology}

In disciplines such as \textit{information systems} (IS) research or BPM research, theoretical and technical, i.e.\ practical, aspects interlink. Here, research is often conducted in an applied manner, in order to solve certain problems. A widely accepted paradigm that aims at enabling explicitly applicable research results is \textit{design science research} (DSR)~\citep{peffersDesignScienceResearch2007}.

\subsection{Design Science Research}
\label{sec:dsr}

\begin{figure}[H]
\centering
\includegraphics[width=\textwidth]{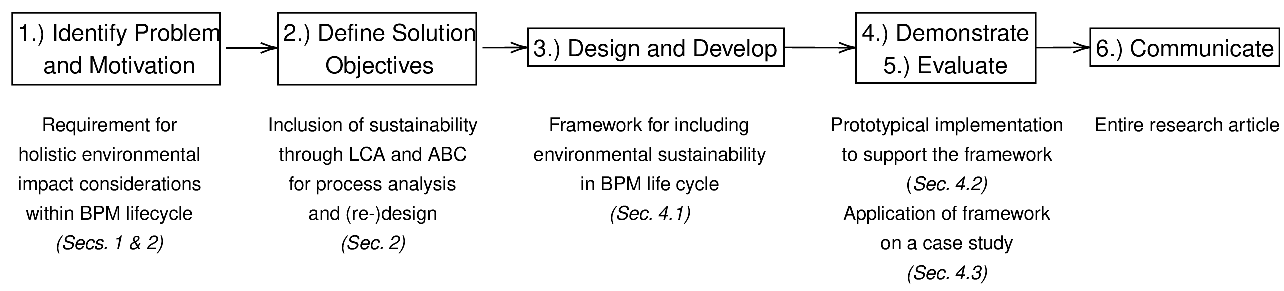}
\caption{Methodology of design science and its application in this work, adapted from \cite{peffersDesignScienceResearch2007}}
\label{fig:dsr}
\end{figure}

DSR, as outlined by \cite{peffersDesignScienceResearch2007} and \cite{wieringaDesignScienceMethodology2014}, is a methodology in IS research that is primarily concerned with solving problems through the creation and evaluation of artifacts. The fundamental principle is the acquisition of knowledge and understanding of a problem through these artifacts~\citep{reckerScientificResearchInformation2013}. The notion of artifact in DSR contains anything that is constructed by humans (i.e., artificial) --- be it a methodology, a conceptual framework, or a specific software. DSR describes a process, with which observed problems can be addressed, research contributions can be provided, designed solutions can be evaluated, and the results can be communicated~\citep{peffersDesignScienceResearch2007}. Figure~\ref{fig:dsr} outlines the process of DSR and how it is applied in this work. Following \cite[Ch. 18]{wieringaDesignScienceMethodology2014}, SOPA is evaluated in a specific context to investigate its implications, through a \emph{single-case mechanism experiment}. Single-case mechanism experiments are frequently used in BPM-related research where software-based artifacts are created (see, e.g., \cite{borkowskiFailure2019} and \cite{klinkmulerDebugging2021}), since they allow us to investigate and evaluate implementations in real-world contexts (either in simulated or in field settings) by observing inputs and outputs of an artifact in response to an experimental treatment \citep[Ch. 5]{wieringaDesignScienceMethodology2014}. Through this, we can test and better understand the mechanisms --- here, being SOPA --- behind a technical prototype \citep{wieringaDesignScienceMethodology2014}. We show the potential of SOPA with an extended example and demonstration. In our case, in accordance with the concept of single-case mechanism experiments, this can be considered as an experimental evaluation in order to assess the overall approach and its potential benefit. %

As the first step of DSR, where the problem domain is investigated and the research itself is motivated, this work argues that, given the pressing matter of climate change and a need for environmentally sustainable business processes, a need for SOPA is underlined. Sections~\ref{sec:intro} and~\ref{sec:background-related} provide a thorough discussion of this.
Solution objectives are further considered in \autoref{sec:background}, where a need to holistically integrate environmental sustainability impact assessment into the BPM life cycle is identified through a literature search and related literature reviews. Based on these objectives, the framework SOPA, which uses LCA and ABC, is designed through a formal-deductive approach \citep{meredithAlternativeResearchParadigms1989} and described in detail in \autoref{sec:results-frame}. In a first iteration, the concepts of ABC and LCA were formalized in line with BPM notions, to derive a theoretical formal framework. This was implemented prototypically and applied to a small set of synthetic use cases, after which the formalisms of the framework and their prototypical implementation were improved. After presenting the result to a number of research and practitioners in the areas of BPM and sustainability, the formalisms, the framework, and the prototypical implementation were finalized in a third iteration. Finally, corresponding to the 4th and 5th step of DSR, SOPA is demonstrated and evaluated by means of an implementation (described in Sec.~\ref{sec:results-system}) on a real-world case study in \autoref{sec:results-eval}. The entire research article represents the 6th step of DSR, where the problem, its importance, and the resulting artifact are presented to transmit the knowledge gained during the DSR process. How data was elicited for this case study is further illustrated in \autoref{sec:data}. The methodology used for designing this implementation is described in more detail in \autoref{sec:sys-design}.

\subsection{System Development}
\label{sec:sys-design}

\begin{figure}[H]
\centering
\includegraphics[width=0.7\textwidth]{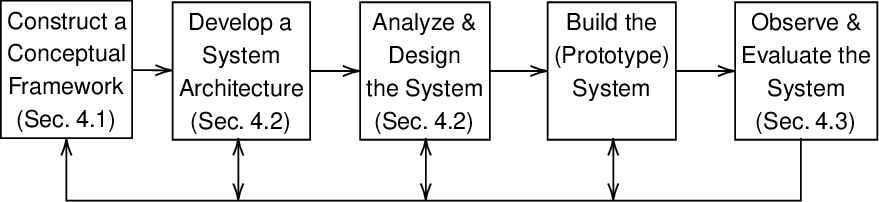}
\caption{System development research methodology, adapted from \cite{nunamakerSystemsDevelopmentInformation1990}}
\label{fig:systems-development}
\end{figure}

While DSR is used to design SOPA, with SOPA being the resulting artifact of the DSR methodology, a concrete software system is also required to facilitate the evaluation and demonstration of SOPA, and to support the overall approach in a practical setting. Here, a system development methodology is followed, where, based on SOPA as a conceptual framework, a system architecture is developed, the system is analyzed and designed, and finally built and evaluated~\citep{nunamakerSystemsDevelopmentInformation1990}. Figure~\ref{fig:systems-development} shows the system development process and its steps.

The system functionalities and requirements to be developed, are implicitly part of the SOPA framework designed with DSR (see \autoref{sec:dsr}), which the system aims to implement on a technical level. Based on these requirements, a concrete system architecture with specific components that have specific functionalities is outlined in \autoref{sec:results-system}. This includes components such as a concrete BPS engine, and a tool for analyzing and calculating impact measures. Further, the process used to carry out the respective system functions (i.e., to conduct sustainability-oriented process analysis and re-design), is illustrated in \autoref{sec:results-system}. Finally, the system is built and evaluated with software components developed specifically for this purpose in \autoref{sec:results-eval}. These resulting components are available as open-source software\footnote{See \url{https://github.com/INSM-TUM/sustainability-analysis-tool} and \url{https://github.com/INSM-TUM/sustainability-scylla-extension} [Accessed: July 17, 2024]}. Thus, the evaluation of the concrete system also serves as a general evaluation of SOPA, following the penultimate step of DSR, by demonstrating its applicability and utility in a practical setting. 

\subsection{Case Study --- Data Collection}
\label{sec:data}

In this work, SOPA is evaluated through application with a case study. For this, a hiring process in a university setting is assessed, analyzed, and investigated for potential to reduce its environmental impact. Hiring processes, which take place in many organizations, are considered to be of high importance and the most costly and time-intensive human resource processes \citep{munstermannPerformanceImpactBusiness2010}. Especially in German academic institutions, hiring processes are executed frequently due to a high degree of personnel fluctuation.\footnote{For example, the TU Berlin had more than 1200 \textit{new} staff appointments in the year of 2021, see \url{https://www.static.tu.berlin/fileadmin/www/10000000/Ueber_die_TU_Berlin/Organisation/100-Tage-Programm/Willkommenstag.pdf} [Accessed: July 12, 2024]} The relevant data has been elicited via unstructured, qualitative stakeholder interviews and document analysis. Two chairs of a department and their assistant were interviewed, firstly to elicit a process model of the hiring process, and second, to identify factors relevant to both the LCA analysis and the process simulation.
Further, the interviews and other relevant documentation (such as a comprehensive guide on required documents) were used to reach reasonable estimates or assumptions, in case no sufficient information could be elicited.

\begin{figure}[H]
    \centering
    \includegraphics[width=\textwidth]{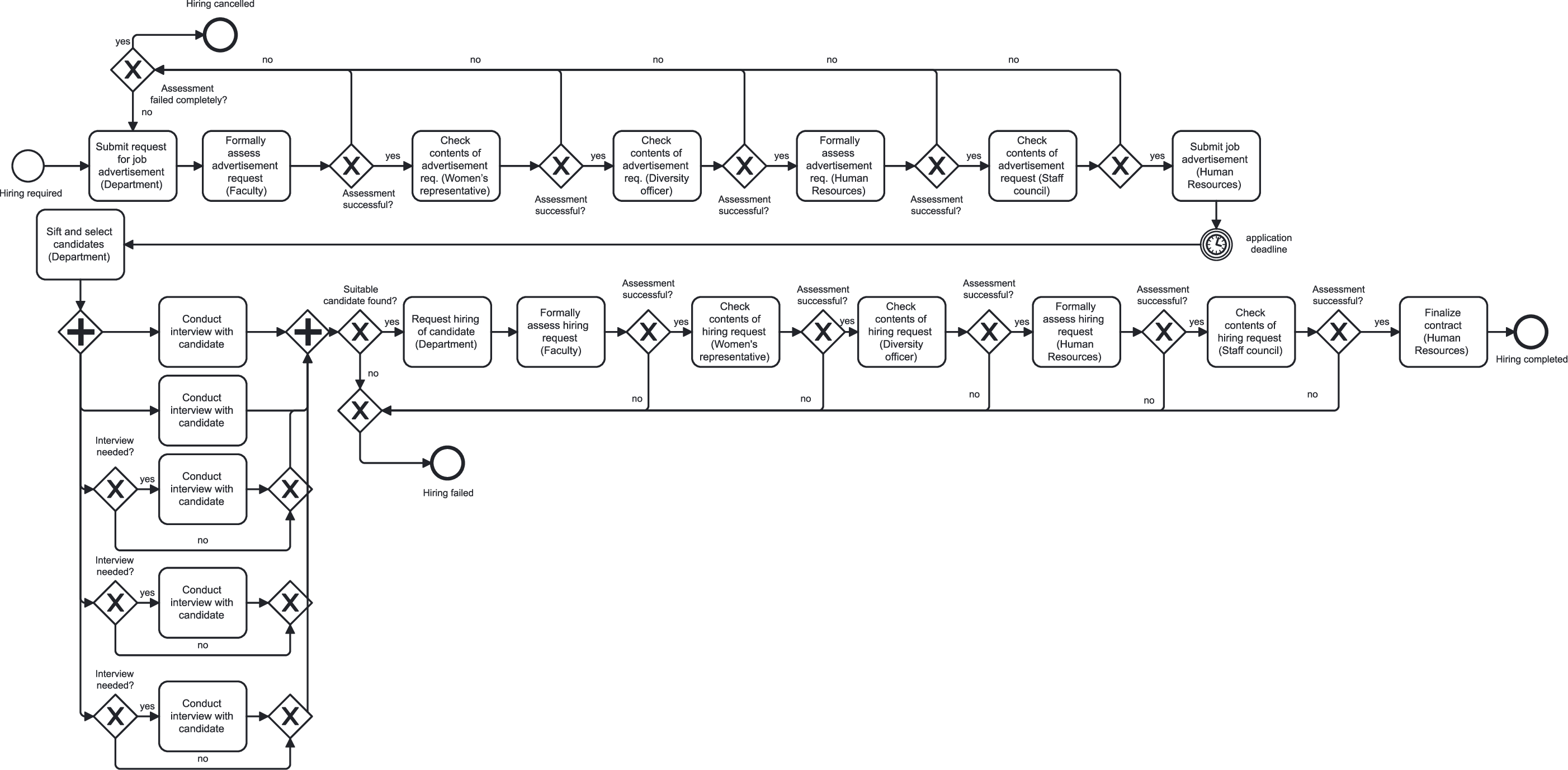}
    \caption{BPMN diagram displaying the hiring process elicited through stakeholder interviews}
    \label{fig:hiring_process_no_driver}
\end{figure}

The LCA analysis was conducted with \textit{mobius}\footnote{See \url{https://mobius.ecochain.com/} [Accessed: July 17, 2024]} a web-based tool for LCA analyses. The concrete LCA method used was \textit{EF 3.0 (Ecoinvent v 3.8 Cut-Off)}~\citep{europeancommissionSupportingInformationCharacterisation2018}, with normalization and weighting (see \autoref{sec:background-lca}). Multiple impact scores were calculated in reliance on the Ecoinvent database v3.8\footnote{See \url{https://ecoinvent.org/the-ecoinvent-database/data-releases/ecoinvent-3-8/} [Accessed: July. 17, 2024]}.
The data elicited for this case study is available online\footnote{See \url{https://figshare.com/s/b0837223b9bf5e109859} [Accessed: July 17, 2024]}. Figure~\ref{fig:hiring_process_no_driver} displays that process, which will be used subsequently as a running example to explain the application of SOPA. A higher resolution of the process model is available online at the same location.

\paragraph{Hiring process.}

The process starts with the need for hiring of an employee in a department. An official request for a job advertisement is posted and checked content-wise and formally by various stakeholders (such as the faculty or the diversity officer). Should one of the assessments fail, a new request can be submitted, or the hiring is cancelled. If all checks have been successful, the job advertisement is submitted to the HR department. Afterward, once the application deadline has elapsed, the candidates are sifted and selected. Subsequently, the selected candidates are interviewed. Due to local gender equality laws, at least two candidates of differing gender are interviewed. In practice, however, we find that usually no more than five candidates are considered potential applicants. After the interviews, if a suitable candidate is found, their employment is requested. Once more, various stakeholders check the request. Should no suitable candidate exist after the interviews, or should the final checks fail, the hiring process has failed. If the final checks are passed, the contract is finalized with the HR department and the hiring is completed. 

\section{Results: SOPA Artifact Design and Evaluation}
\label{sec:results}

Following the DSR methodology as described above, the results of this study are the design, implementation, and evaluation of our artifact, the SOPA framework. 
Accordingly, in this section we first present SOPA, including its application procedure and the resulting metamodel.
Then, we describe our implementation of the framework in a prototype system, which enables the evaluation based on a case study, covered in the last part of this section.

\subsection{Framework and Metamodel}
\label{sec:results-frame}

\begin{figure}[H]
\includegraphics[width=\textwidth]{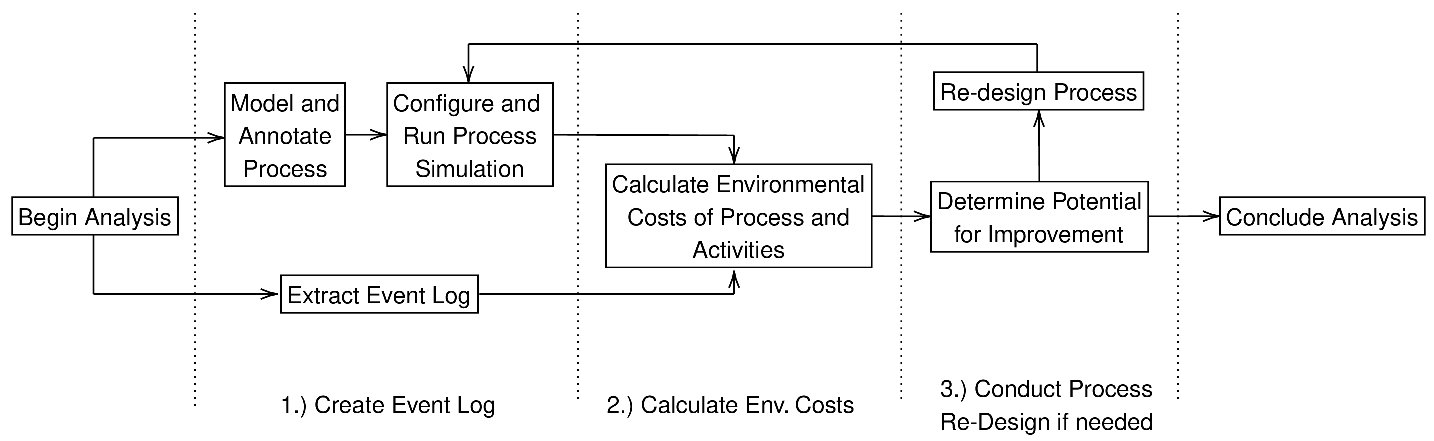}
\caption{Application procedure of SOPA}
\label{fig:application_procedure}
\end{figure}

SOPA proposes a three-step approach for analyzing business processes for their environmental impact and their sustainability-oriented re-design. When applying SOPA to analyze business processes for their environmental impact and to re-design them accordingly, two possible scenarios exist. Either an existing event log, which contains execution data of real-world process instances, is analyzed and used for reasoning about improvement potentials (\textit{log-based analysis}), or a process model is, through usage of BPS, assessed (\textit{model-based analysis}). This leads to the following three central steps of SOPA:
\medskip
\begin{enumerate}
    \item Either, a process model is created and annotated, and simulated to generate synthetic execution data for further assessment. Alternatively, an event log containing execution data of real-world process instances, needs to be extracted from the corresponding information system and provided.
    \item The execution data is used in conjunction with LCA data to calculate the environmental costs of the process and its activities.
    \item Should a need for process re-design become apparent, process simulation can be used further to investigate potential improvements.
\end{enumerate}
\medskip
\noindent
Figure~\ref{fig:application_procedure} displays the overall application procedure, with the boxes representing the concrete tasks, separated along the three central steps of SOPA as described above. Following the arrows, after deciding to begin the analysis, either the model-based analysis on the upper path, or the model-based analysis on the lower path is conducted. Re-design iterations, which are conceptually model-based, lead to another iteration of the model-based path. Otherwise, the analysis is complete, and process experts can deliberate the findings. Notably, SOPA works without simulation, given that the required LCA data is present in the event log. This is a strong requirement for the evaluation, however, therefore this work applies process simulation to illustrate the application in terms of analysis and re-design.

In the following, \autoref{sec:results-frame-modelling-sim} starts with presenting the modeling- and simulation-related activities and concepts of SOPA, and requirements towards extracted event logs. \autoref{sec:results-frame-calc} details the environmental cost calculation and concepts, while \autoref{sec:results-frame-re} illustrates the steps for sustainability-oriented process re-design of SOPA. We present SOPA using two types of formal representation: On the one hand, metamodels provide a high-level overview of the entities and user-provided/emergent data that are required by SOPA. On the other, algebraic specifications define the relationships of the metamodel more strictly, and provide an understanding of how the actual calculations underlying SOPA can be implemented. Moreover, we provide illustrating examples in \autoref{sec:appendix0}, which serve to detail an example instance of a prototypical implementation and application of a SOPA-based analysis.

\subsubsection{Event Log Simulation and Extraction}
\label{sec:results-frame-modelling-sim}

\begin{figure}[H]
    \centering
    \includegraphics[width=0.5\textwidth]{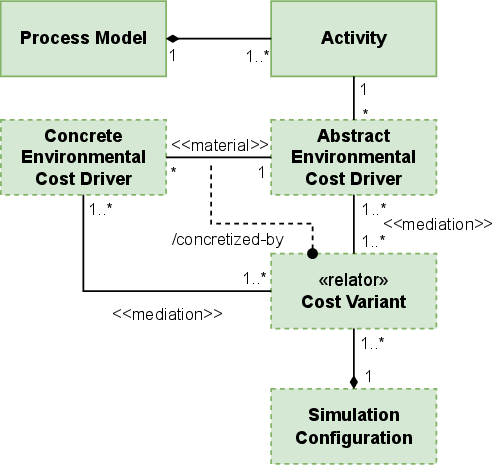}
    \caption{Metamodel of modeling-related concepts of SOPA. Extensions conceptualized in this work are shown with dashed borders}
    \label{fig:meta-model}
\end{figure}

In the following, the model-based variant of the first step of SOPA is explained because it illustrates all modeling concepts required by SOPA. However, requirements towards extracted event logs to be used with SOPA are provided at the end of this section.
Figure~\ref{fig:meta-model} displays a metamodel of these modeling concepts and their relations. Besides this high-level metamodel, we also provide formal definitions of the concepts to guide implementations based on the SOPA framework.

First, as the basis for a model-based analysis, the process needs to be modelled, e.g.\ by using process modeling languages such as BPMN~\citep{omg2011bpmn}\footnote{While we use BPMN for process modeling due to its industry-wide adoption as a standard, prevalent tool support, and extensibility \citep{pufahlBPMNHealthcareChallenges2022}, SOPA conceptually allows different modeling languages to be used, as long as all concepts formalized in this work can sufficiently be expressed.}. Activities form the basis of processes and process models, as shown in Fig.~\ref{fig:meta-model}. Definition~\ref{def:activity} expresses this formally.

\medskip
\begin{Definition}{(Activity)}
\label{def:activity}
Let $A$ be the universe of all possible activities. Then, $a\in A$ is a single activity that can be enacted during process execution.
\end{Definition}
\medskip

Afterward, \textit{abstract environmental cost drivers} are annotated to the individual activities. They describe, on an abstract level, what objects, resources and products, i.e., everything that is generally assessable through LCA, are involved in the execution of specific activities.
On the other hand, \textit{concrete environmental cost drivers} describe the environmental impact of concrete instantiations of their abstract counterparts, based on LCA analyses through individual LCA scores. For example, the specific mode of transport used by the in-house mail could be either a petrol-powered car or an electric bicycle, leading to different impact scores for the two possible variants. Figure~\ref{fig:hiring_process_driver} shows a process model, subsequently used for our case study, that has been annotated with abstract cost drivers.
During modeling, abstract and concrete environmental cost drivers are expected to be identified in conjunction (either, abstract or concrete environmental cost drivers are known at first), and continuously refined through abstraction, respective concretization.

The relation between abstract and concrete environmental cost drivers can be organized into a strict hierarchy. This \textit{environmental cost driver hierarchy}, characterized in Definition~\ref{def:hierarchy}, organizes abstract environmental cost drivers and concrete environmental cost drivers, and describes how abstract environmental cost drivers can be concretized during process execution. This represents the materialized relation between abstract and concrete environmental cost drivers shown in the metamodel of Fig.~\ref{fig:meta-model}.

\medskip
\begin{Definition}{(Environmental Cost Driver Hierarchy)}
\label{def:hierarchy}
Let $D$ be a set of abstract environmental cost drivers, where $d \in D$ is a single abstract environmental cost driver. Let $C$ be a set of concrete environmental cost drivers, where $c \in C$ is a single concrete environmental cost driver. Finally, let $H$ be a hierarchy of abstract and concrete environmental cost drivers, so that $H \subseteq D \times C$. This means that abstract environmental cost drivers $d \in D$ has one or more concrete environmental cost drivers $c \in C$ associated with it.
\end{Definition}
\medskip
\noindent
For each concrete environmental cost driver, a concrete, LCA-defined impact score, or cost, can be determined --- either, through manual LCA analyses, or through an information system used for process execution, that already supplies this information. This is realized through a cost function described in Definition~\ref{def:cost-fn}, where for each concrete environmental cost driver, the impact score is returned.

\medskip
\begin{Definition}{(Cost Function)}
\label{def:cost-fn}
Let $C$ be a set of concrete environmental cost drivers. %
Cost functions assign concrete environmental costs in terms of impact on sustainability to concrete environmental cost drivers. More specifically, let
$cost$ be a cost function of concrete environmental cost drivers so that $cost: C \rightarrow \mathbb{Q}$ assigns environmental cost values to concrete environmental cost drivers. %
\end{Definition}
\medskip
\noindent

An illustrating example of an environmental cost driver hierarchy and a cost function is provided in \autoref{sec:example-hierarchy-cost}.
The LCA scores assigned to concrete environmental cost drivers need to have either been part of the recorded event data, or be elicited through LCA analyses of process experts with an LCA method that aggregates the LCA results into a single score. Notably, the quality of the SOPA-led analysis is determined by how precisely the event log has captured the process executions, respective represents them accurately. Further, the LCA analyses and their exactness determine how exact the impact of concrete environmental cost drivers, and thus, the overall process, can be assessed. Here, making use of LCA databases (see \autoref{sec:background-lca}) which could be shared industry-wide to disseminate accurate data for certain materials, products, or services, might increase the analysis' accuracy and reduce the time needed for conducting the LCA analyses.

\textit{Cost variants} govern what specific combinations of concretizations can occur during individual process instances based on the environmental cost driver hierarchy --- in other words, what sets of concrete environmental cost drivers can, during process execution, take the place of the abstract environmental cost drivers during activity execution. For example, one concrete mode of transport (i.e., via car or bike) requires the documents to be printed (represented with one concrete environmental cost driver), while another mode (i.e., sending them via email instead) requires them to exist as digital documents (represented with a different concrete environmental cost driver). In Fig.~\ref{fig:meta-model}, this is shown by the mediation of concrete and abstract environmental cost drivers through the cost variant.
Further, cost variants can be used to emulate cost distributions: when, e.g., one abstract environmental cost driver is expected to occur with a continuous distribution, this distribution can be divided into several ranges, for which concrete LCA scores and thus specific cost variants can be determined. Based on the expected frequency of the respective ranges, their cost variants can be simulated, and thus, the distribution be approximated.

Finally, the overall process simulation can be configured through a simulation configuration (see \autoref{sec:background-simulation}), by not only selecting cost variants and their frequency of occurrence, but also by determining conventional parameters of process simulation, such as gateway probabilities and event arrival rates.
The concepts outlined in this section constitute a basis for process simulation, through which a simulated event log can be generated, that can serve as the basis for the second step of the framework (i.e., the calculation of the environmental costs, see Fig.~\ref{fig:application_procedure}).%

The individual cost of concrete environmental cost drivers in relation to cost variants can be modelled or provided with a cost variant config, such as the illustrating example shown in Listing~\ref{lst:cost-variant-excerpt} in \autoref{sec:example-cost-variant}. As described above, they identify the specific combinations of concretizations that can occur during process instances. Further, the environmental cost for each concretization of the environmental cost drivers is provided here, representing the cost function described above in Definition~\ref{def:cost-fn}. Notably, such a configuration file can also be used to concretize environmental cost drivers in event logs that have been extracted out of an IS and not just simulated, provided information about the variants and abstract environmental cost drivers to be concretized is present.

\subsubsection{Environmental Cost Calculation}
\label{sec:results-frame-calc}

In the second step of SOPA (see Fig.~\ref{fig:application_procedure}), the environmental cost of business processes is calculated based on event logs, which represent recorded executions through information systems, in a manner that is similar to existing ABC approaches. The objective here is, to gain insights into the environmental impact of a process by considering its recorded process and activity instances.

\begin{figure}[H]
    \centering
    \includegraphics{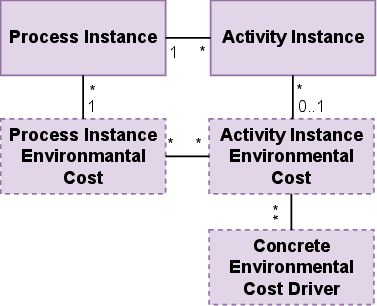}
    \caption{Metamodel of instance-related concepts of SOPA. Extensions conceptualized in this work are shown with dashed borders}
    \label{fig:meta-instance}
\end{figure}

Figure~\ref{fig:meta-instance} outlines how environmental costs of activity and process instances and concrete environmental cost drivers relate. Their interrelation is the basis for the environmental cost calculation of SOPA, which is more formally described by the definitions this section subsequently provides.

As discussed above, in the case of a model-based analysis, concrete LCA scores for the concrete environmental cost drivers assigned to activity instances during simulation configuration need to be provided. In the case of a log-based analysis, sufficient information needs to be present in the log, providing either concrete LCA scores for the activities, or information about abstract environmental cost drivers and how they are concretized.%

Having an event log that captures process executions, and LCA scores expressing the environmental impact of individual environmental cost drivers involved in activity execution, an ABC analysis can be conducted, to assess the overall environmental impact of the process. For each instance of an activity or a process, the respective cost (i.e., its environmental impact) can be calculated. %
The following provides a formal definition of the concepts, and how impact scores can be calculated.

As shown in the illustrating example provided in Listing~\ref{lst:xes-excerpt} of \autoref{sec:example-event-log}, event logs contain recorded instances of processes, consisting of recorded activity executions. In the case of SOPA, activities are instantiated and executed during process execution, and each activity instance is potentially associated with certain concrete environmental cost drivers. The overall possible set of activity instances is determined by the combination of activities and sets of concrete environmental cost drivers, as expressed in Definition~\ref{def:activity-instance}. 
In theory, the event log provided there could either have been extracted from an information system, or created synthetically using process simulation. In both cases, activity instances (i.e., events) contain information about environmental cost drivers, and process instances (i.e., traces), information about the respective cost variant. For an illustrating example of how such an event log might look like, we refer to Listing~\ref{lst:xes-excerpt}. Either, by being recorded in the information system, or by being modelled during process simulation, knowledge of the environmental cost driver and cost variants contained in recorded events and traces can be used to compute the overall environmental impact of the business process for which the log is available.

\medskip
\begin{Definition}{(Activity Instance)}
\label{def:activity-instance}
Let the set of activity instances $I$ be a set $I \subset A \times \mathcal{P} (C)$, where each activity is associated with a set of concrete environmental cost drivers. A single activity instance $i \in I$ therefore is a pair $(a, Q)$, with $a \in A,\ Q \in \mathcal{P}(C)$.
\end{Definition}
\medskip
\noindent
The assignment of concrete environmental cost drivers to an activity instance is either done during process execution and recorded in an event log, or done during simulation time --- here, as described in the previous section, the simulation configuration, cost variant, and environmental cost driver hierarchy govern how activities in the model and the annotated abstract environmental cost drivers are translated to activity instances and concrete environmental cost drivers.
For the simulation of a single process instance, the cost variant is determined by the simulation configuration, which in turn describes for each activity, how the annotated abstract environmental cost drivers are turned into concrete ones. Since this is more of a technical challenge than a conceptual problem, this translation is not formalized further. Conversely, during actual execution, the concrete environmental cost drivers  can directly be recorded for all activity instances of the process instance.

This means that, following Definition~\ref{def:process-instance}, process instances are sequences of activity instances (see also the relation between activity instances and process instances in Fig.~\ref{fig:meta-instance}), and they either emerge during execution or simulation of the process.

\medskip
\begin{Definition}{(Process Instance)}
\label{def:process-instance}
A process instance is a finite non-empty sequence of activity instances, containing totally ordered pairs of activity instances and sets of concrete environmental cost drivers. Let $I \subset A \times \mathcal{P} (C)$ be an alphabet of activity instances. Then, $I^*$ is the set of all finite sequences of activity instances, which we also call process instances.
Thus, let $t \in I^*$ be a single process instance, which contains one or more pairs $(a, Q)$.
\end{Definition}
\medskip
\noindent
For an illustrating example of process instances, see \autoref{sec:example-process-instance}. Note that we rely on a formalization of sequences provided by~\cite{vanderaalstProcessMining2016} that assumes that the sequence's length can be known, elements can be retrieved, and that sequences can contain the same element multiple times with an explicit order, which we omit here for clarity. Moreover, we assume a total order of all activity instances belonging to a process instance in line with established formalizations of process instances. This allows i.a., process mining algorithms to be applied to a set of process instances, which commonly rely on totally ordered activity instances, see e.g.\ \cite{vanderaalstProcessMining2016}.

For each activity instance, the overall environmental impact can be calculated, by considering all concrete environmental cost drivers of that instance. Definition~\ref{def:activity-instance-cost} describes this calculation, for which we refer to \autoref{sec:example-activity-instance-cost} for an illustrating example.

\medskip
\begin{Definition}{(Environmental Activity Instance Cost)}
\label{def:activity-instance-cost}
Let $(a, Q) \in I$, consisting of an activity $a \in A$ and a set of concrete environmental cost drivers $Q \subset \mathcal{P} (C)$, be an activity instance. For each instance, the environmental cost can be calculated with a function $\mi{activity\_instance\_cost}: I \rightarrow \mathbb{Q}$ so that:

$\mi{activity\_instance\_cost}(a, Q) = (\sum\limits_{q \in Q} cost(q))\ $ %
\end{Definition}
\medskip
\noindent
Based on activity instance costs, the cost of a process instance can similarly be calculated by summing up the costs of the involved activity instances, following Definition~\ref{def:process-instance-cost}. We refer to \autoref{sec:example-process-instance-cost} for an illustrating example.

\medskip
\begin{Definition}{(Environmental Process Instance Cost)}
\label{def:process-instance-cost}
Let $T \subset I^*$ be a set of process instances. Then, for a single process instance $t \in T$, the environmental cost can be calculated with a function $\mi{process\_instance\_cost}: T \rightarrow \mathbb{Q}$ so that:

$\mi{process\_instance\_cost(t)} = \sum\limits_{(a, Q) \in t}\mi{activity\_instance\_cost}(a, Q)$
\end{Definition}
\medskip
\noindent
This provides a concrete numerical score of how impactful a single process instance has been. However, not just individual instances of activities and the overall process can thus be considered, but also the \textit{average} environmental instance cost of specific activities, respective the process itself, can be calculated. This is further described in Definitions~\ref{def:average-activity-cost} and~\ref{def:average-process-instance-cost}. Notably, a collection of process instances is formalized as a multiset of process instances -- while traditionally (see e.g.~\cite{vanderaalstProcessMining2016}), they are formalized differently, to be able to differentiate between instances based on e.g., the case they represent, this differentiation is not needed for the purpose of calculating the environmental impacts, and multiple instances can be, considering just the involved activities and the concrete environmental cost drivers, the same.
For an illustrating example, see \autoref{sec:example-average-activity-instance-cost}.

\medskip
\begin{Definition}{(Average Environmental Activity Cost)}
\label{def:average-activity-cost}
Let $T \in \mathcal{B}(I^*)$ be a multiset of process instances, and $a \in A$ be an activity. Let $\mi{occurrence\_count}: (A \times T) \rightarrow \mathbb{N}$ be a function that counts the occurrences of $a$ in a sequence of activity instances $t \in T$. Let $\mi{specific\_count}: (I \times T) \rightarrow \mathbb{N}$ be a function that counts the occurrences of activity instances $a$ with a specific set of concrete environmental cost drivers $Q$ in a sequence of activity instances $t \in T$.
Then, the average environmental cost of a specific activity across all process instances can be calculated, and $\mi{average\_activity\_cost}: A \times \mathcal{P}(T) \rightarrow \mathbb{Q}$ can be defined so that:

$\mi{average\_activity\_cost}(a, T) = \frac{\sum\limits_{Q \in \mathcal{P}(C)} \sum\limits_{t \in T} \mi{specific\_count}(a,Q, t) \cdot \mi{activity\_instance\_cost}(a,Q)}{\sum\limits_{t \in T} \mi{occurrence\_count}(a, t)}$
\end{Definition}
\medskip

\noindent
The average environmental cost of all process instances can be calculated similarly, as described in Definition \ref{def:average-process-instance-cost}. See \autoref{sec:example-average-process-instance-cost} for an illustrating example.

\medskip
\begin{Definition}{(Average Environmental Process Instance Cost)}
\label{def:average-process-instance-cost}
Let $T \in \mathcal{B}(I^*)$ be a multiset of process instances. For this $T$, the average environmental cost can be calculated with $\mi{average\_process\_instance\_cost}: \mathcal{P}(T) \rightarrow \mathbb{Q}$ so that:

$\mi{average\_process\_instance\_cost}(T) = \frac{\sum\limits_{t \in T}\mi{process\_instance\_cost}(t)}{|T|}$
\end{Definition}
\medskip

\noindent
These calculated costs, whose conceptual relation is outlined in Fig.~\ref{fig:meta-instance}, then provide a notion of how costly activity instances, respective process instances, can be expected to be, based on all recorded process executions and the cost information they provide.

Using these costs, potential for improvement can be uncovered, and the activity of process re-design as further described in the subsequent \autoref{sec:results-frame-re} can be conducted.

\subsubsection{Process Re-Design}
\label{sec:results-frame-re}

The third and final step of SOPA, as shown in Fig.~\ref{fig:application_procedure}, concerns the sustainability-oriented process re-design. When a consideration of the calculated environmental casts uncovers potential for improvement (e.g., disproportionally high average environmental costs of process instances, specific average environmental costs of activities, or a reduction not being as high as expected), the process can be re-designed accordingly and simulated to test hypotheses on whether the re-design could actually improve, i.e., reduce, the environmental impact of the process under consideration.

\begin{figure}[H]
    \centering
    \includegraphics[width=0.9\textwidth]{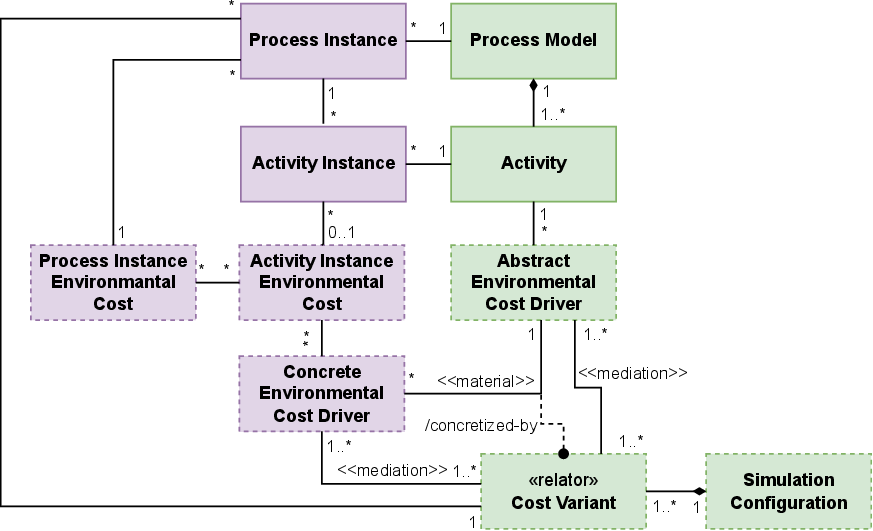}
    \caption{Class diagram showing the entire metamodel of SOPA. Extensions conceptualized in this work are shown with dashed borders. Entities in green are provided during design time, entities in purple either emerge based on probabilistic distribution during simulation time, or are instantiated during process execution. Since concrete environmental cost drivers can both be modeled and be part of activity instances, they are shown in both colors}
    \label{fig:meta}
\end{figure}

Figure~\ref{fig:meta} shows the interplay between instance- and model-related concepts in its entirety. While the formal definitions above have underlined how activity and environmental process costs can be calculated, based both on instance and modeled data, the metamodel of Fig.~\ref{fig:meta} provides a higher-level overview of the information and entities that a required for SOPA, and how they are combined for the formally defined calculations.

Using these model- and instance-related concepts, SOPA can be used to conduct business process re-design across three facets:
\smallskip
\begin{enumerate}
    \item \textit{Concrete environmental cost drivers} can be changed, which represents an adaption of the respective activities to use, for example, less impactful materials.
    \item The \textit{simulated behavior} can be altered (e.g., gateway probabilities), representing a change in process execution, facilitated i.a. through changes in the processes operational support.
    \item Both \textit{environmental cost drivers} and \textit{simulated behavior} can be changed \textit{simultaneously}, implying a causal relation between a new concrete environmental cost driver and the behavior. This could, e.g., be the case when the adaption of a process to use less impactful materials in one activity leads to a higher probability of having to execute another activity. An example could be a logistics process, where using structurally weaker packaging material that is less impactful can lead to a higher probability of needing to handle return shipments due to breakage, in turn leading to an overall increase in impact.
\end{enumerate}

\smallskip
\noindent
After configuring a new simulation run, the resulting event log can be analyzed once again, and the cost analyses can be compared. Either, a new re-design iteration may be necessary, or the theoretical re-design can be implemented in practice. Further, the process analyst could consider whether the potential reduction in environmental impact may be too costly to implement, and further tweak the re-design scenario to take economic considerations into account.

\subsection{System Development}
\label{sec:results-system}

To support the approach of SOPA, and to facilitate the case study-based evaluation, a prototypical software system has been developed. This software system realizes the overall metamodel, as displayed in Fig.~\ref{fig:meta}, and the formalisms of SOPA, and allows for a practical application of SOPA's application procedure of Fig.~\ref{fig:application_procedure}. Note that the software system is not the central artifact, but rather a means for evaluating it. The following briefly presents the respective components, which combine modeling- and instance-related concepts. 

\begin{figure}[H]
    \centering
    \includegraphics[width=0.5\textwidth]{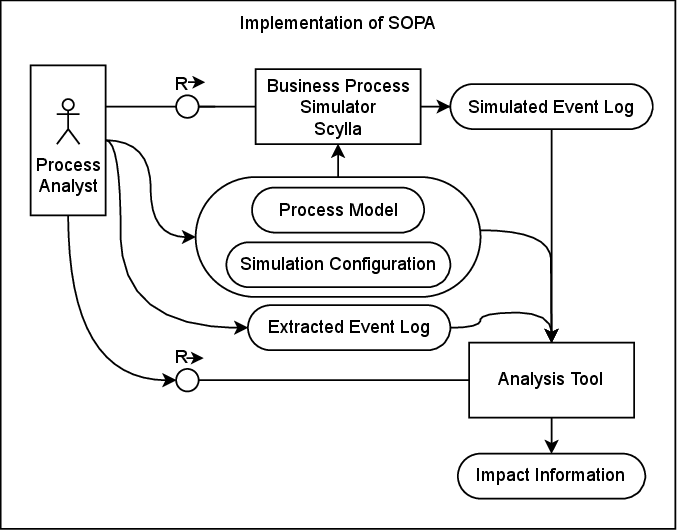}
    \caption{Fundamental Modeling Concepts (FMC)~\citep{knopfelFundamentalModelingConcepts2005} diagram displaying the concrete architecture of the implementation supporting SOPA. FMC differentiates between active and passive components. Active components such as the Analysis Tool or the Process Analyst, shown in square, perform activities in the system. Passive components such as the Process Model, displayed in oval, store, contain, or transmit information}
    \label{fig:concrete_architecture}
\end{figure}

The system's architecture consists of multiple components used in concert by a process analyst to elicit environmental impact information, consisting of average process instance and activity environmental costs, of a business process under investigation. Figure~\ref{fig:concrete_architecture} illustrates the different components that are involved.
First, the process analyst can provide a process model annotated with abstract environmental cost driver information and a simulation configuration with cost variants to a fork of the process simulation engine \textit{Scylla}\footnote{See \url{https://github.com/INSM-TUM/sustainability-scylla-extension} [Accessed: July 17, 2024]} (see~\cite{pufahlDesignExtensibleBPMN2018}), that has been modified to accommodate the SOPA-specific concepts. 
The resulting (simulated) event log is used by the \textit{Analysis Tool}\footnote{See \url{https://github.com/INSM-TUM/sustainability-analysis-tool} [Accessed: July 17, 2024]}, together with the simulation configuration (containing cost variants and the corresponding concrete environmental cost drivers) and process model provided to Scylla, to calculate the relevant impact information. Alternatively, an event log containing the relevant information that has been extracted out of a real-world information system could be used as well. The impact information can then be used by the process analyst to either put the hypothesized changes into practice, or to drive a new re-design iteration.

\subsection{Case Study Evaluation}
\label{sec:results-eval}

In the remainder of this section, SOPA is evaluated by being applied to the university hiring process case study that is depicted in Fig.~\ref{fig:hiring_process_no_driver} through the system outlined above. This constitutes a \textit{single-case mechanism experiment}~\citep[Ch. 18]{wieringaDesignScienceMethodology2014}. Concretely, we aim to address the following \textit{evaluation goals}:
\smallskip
\begin{enumerate}
    \item Show that the SOPA implementation allows a quantification of the environmental impact of activities and process instances
    \item Show that SOPA can be used to assess the hypothetical reduction of environmental impact incurred by re-designing a business process in various ways
    \item Illustrate the holistic nature with which SOPA captures environmental impact
\end{enumerate}
\smallskip
\noindent
Since the hiring process is largely executed manually with neither an IT system nor an IT system that logs environmental cost driver information, the model-based variant of SOPA is being used here. The process model itself, as well as the environmental cost drivers and the relevant LCA data have been elicited through stakeholder interviews and use of LCA tooling and databases, as described in \autoref{sec:data}.

\paragraph{Environmental Cost Drivers and Impact Data.}

The following abstract environmental cost drivers, listed in Table~\ref{tab:cost-driver}, have been identified through the stakeholder interviews: %

\begin{table}[h] 
\caption{Abstract environmental cost drivers identified for the hiring process\label{tab:cost-driver}}
\begin{tabularx}{\textwidth}{c|X}
\toprule
\textbf{Abstract environmental cost driver} & \textbf{Description}\\
\midrule
Request for job advertisement & The cost of creating the documents required for requesting a job advertisement.\\\midrule %
In-house mail & The cost of transporting the documents between offices.\\\midrule %
Advertisement & The cost of creating the documents representing the job advertisement.\\\midrule %
Sifting & The cost of sifting through all applications.\\\midrule %
Interview & The cost of conducting a single job interview.\\\midrule %
Application for employment & The cost of creating the documents required for requesting the employment of a specific applicant. \\\midrule %
Contract documents & The cost of creating the documents of the final employment contract.\\ %
\bottomrule
\end{tabularx}
\end{table}

\begin{figure}[H]
    \centering
    \includegraphics[width=\textwidth]{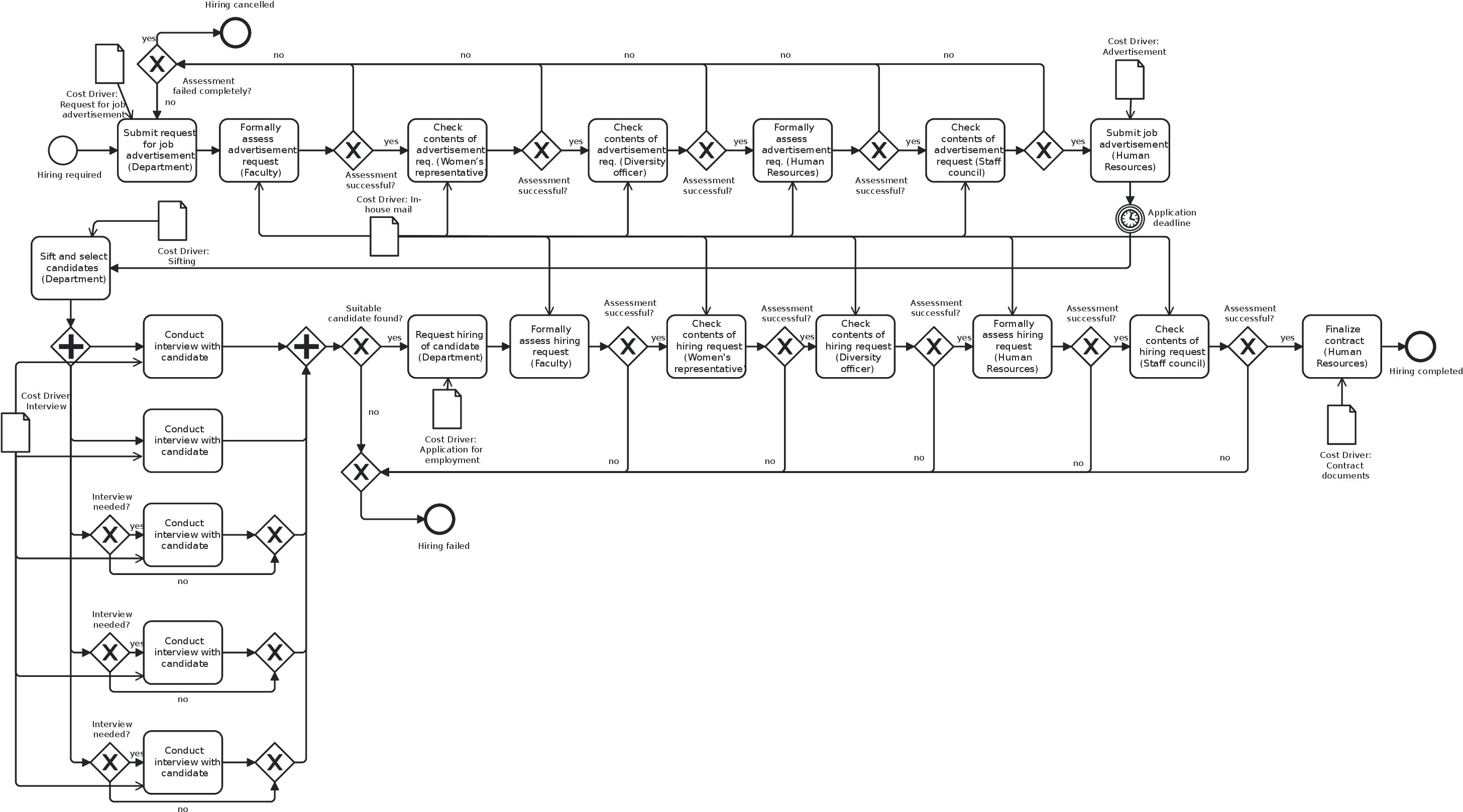}
    \caption{BPMN diagram of the hiring process, with annotated abstract environmental cost driver}
    \label{fig:hiring_process_driver}
\end{figure}

Figure~\ref{fig:hiring_process_driver} displays these abstract environmental cost drivers annotated to the activities of a process model.

\paragraph{Simulation Parameters.}

The following parameters, listed in Table~\ref{tab:sim-params}, have been used for the simulation. Further details can be found in the respective simulation configuration files available online\footnote{See \url{https://figshare.com/s/b0837223b9bf5e109859} [Accessed: July 17, 2024]}. Generally, based on stakeholder interviews, the content-and form related assessments of all requests are assumed to be 95\% successful, while only a probability of 5\% is assumed for cancellation of the entire process. With a probability of 2\%, the process fails due to a lack of suitable candidates. With a probability of 50\% each, three additional interviews may be conducted. Further, the simulation is configured to simulate 500 process instances. Notably, since inter-instance interaction is irrelevant in SOPA, no further configuration of, e.g., instance arrival rates or activity durations is necessary.

\begin{table}[h] 
\caption{Parameters used for process simulation\label{tab:sim-params}}
\begin{tabular}{@{}l|l@{}}
\toprule
\textbf{Parameter}	& \textbf{Probability}\\
\midrule
Probability of successful assessment in terms of content and form & 95\% \\\midrule
Probability of overall cancellation	& 5\% \\\midrule
Probability of each additional interview&50\%  \\\midrule
Probability of failure due to lack of candidates & 2\% \\
\bottomrule
\end{tabular}
\end{table}

\paragraph{Scenarios and Results.}

Three scenarios have been identified and are evaluated SOPA in the following, one representing the \textit{as-is} process, and two representing potential re-designs for environmental impact reduction. Table~\ref{tab:sim-scenarios} describes how the different environmental cost drivers are concretized in the scenario, and provides their respective impact scores elicited with LCA, as described in \autoref{sec:data}.

\textbf{Scenario A} represents the \textit{as-is} procedure: the required \textit{documents} are created digitally (assuming 2h of computer work per document), are printed on paper (assuming 5g per page, the request for advertisement is known to average 8 pages, the advertisement 2 pages, the application for employment 20 pages, and the contract documents 5 pages, printed with a non-specific black-and-white laser printer) and transported between the different offices via \textit{in-house mail} by car (i.e., a 3.5 ton diesel lorry, emission class EUROIII). For all delivery trips, an average distance of 750 m is assumed. The \textit{job interviews} are assumed to be conducted via video conference, for one hour. The \textit{sifting} of documents is assumed to take 6 hours and to be done with the aid of a computer.
\textbf{Scenario B} represents a \textit{hypothetical re-design} scenario, in which the \textit{in-house mail} instead would be transported via an electric bicycle using renewable energy and weighing 20 kg, averaging, as in Scenario A, 750 m per trip. 
Finally, \textbf{Scenario C}  also represents a \textit{hypothetical re-designed} variant of the hiring process where no \textit{documents} would be printed and transported, but instead would be \textit{transmitted digitally}, discounting potential digital long-term storage. File transmission is assumed to take, on average, one minute. Further, the \textit{sifting} of candidates would be less rigorous, taking 3h of computer time instead of 6h. However, due to this less precise sifting, \textit{at least} two of the three additional interviews will be conducted.

\begin{table}[h]
    \caption{Scenarios and respective LCA sores of the concrete environmental cost drivers for evaluation with SOPA\label{tab:sim-scenarios}}
    \begin{tabular}{@{}l||l|l|l@{}}
    \toprule
    \textbf{Environmental cost driver}	& \textbf{Scenario A} & \textbf{Scenario B} & \textbf{Scenario C}\\
    \midrule
    Request for job advertisement & $2.89\times10^{-5}$ & See A. & $1.95\times10^{-5}$ \\\midrule
    In-house mail & $3.91\times10^{-5}$ & $4.22\times10^{-6}$ & $1.51\times10^{-8}$\\\midrule
    Advertisement & $2.19\times10^{-5}$& See A. & $1.95\times10^{-5}$\\\midrule
    Sifting & $5.85\times10^{-5}$ & See A. & $2.93\times10^{-5}$ \\\midrule
    Interview & $3.5\times10^{-5}$ & See A. & See A.  \\\midrule
    Application for employment & $4.31\times10^{-5}$ & See A. &  $1.95\times10^{-5}$\\\midrule
    Contract documents & $2.54\times10^{-5}$ & See A. & $1.95\times10^{-5}$ \\
    \bottomrule
    \end{tabular}
\end{table}

\begin{figure}[htbp]
    \centering
    \includegraphics[width=0.5\textwidth]{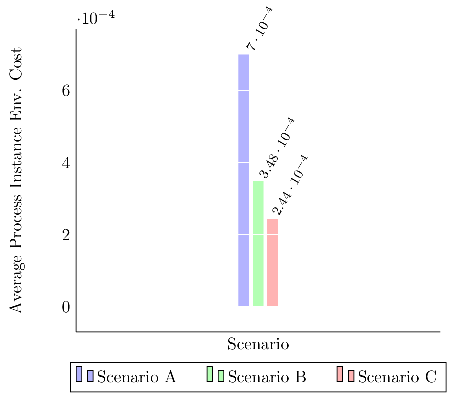}
    \caption{Average environmental process instance costs of the hiring process per scenario, elicited through SOPA}
    \label{fig:results-average-per-scenario}
\end{figure}

Applying SOPA to the hiring process and the three identified scenarios, results in impact information, consisting of average environmental activity costs and average environmental process instance costs, which give a notion of the impact of activity and process executions depending on the scenario. Table~\ref{tab:sim-results} provides a complete overview of the resulting impact information. Notably, the average environmental cost per process instance decreases drastically in the hypothetical re-design scenarios B and C, as shown in Figure~\ref{fig:results-average-per-scenario}.

Switching the mode of transport of the in-house mail would incur an overall impact reduction of $50.18\%$. Additionally, removing all paper-based documents from the process, while reducing the time spent sifting applicants at the cost of having to conduct at least four instead of two interviews, would reduce the environmental impact by $65.08\%$ per process instance, compared to the as-is impact. Further, Fig.~\ref{fig:bar_results} visualizes the average activity scores per scenario, which are also provided in Table~\ref{tab:sim-results} and compared in Table~\ref{tab:sim-results-compare}, all to be found in \autoref{sec:appendix1}. Interestingly, we see that the reduction across almost all cost drivers from Scenario B to C is somewhat compensated by the fact that more interviews will need to be conducted. With this information, process experts and decision makers now can reason about further potential for process impact reduction, by either implementing Scenario B or C, or by further reducing the cost of individual activities, e.g., reducing the maximum duration of the interviews, which constitutes one of the most impactful activity across all two hypothetical re-design scenarios. Additionally, Scenario C could be further revised, since the (economic) cost of having to conduct more interviews and having to implement the infrastructure for paperless applications might not warrant the increase in environmental saving in comparison to Scenario B. Similarly, a combination of Scenarios B and C, so that some process instances are done-paper less and others not, could be explored, to further deliberate the feasibility of the various scenarios.

Notably, this addresses both \textit{evaluation goals}, since we demonstrably are able to quantify the impacts of process instances and activities, and determine how they change across re-design scenarios. We have also illustrated how a consideration of trade-offs between concrete cost drivers and process behavior is possible with SOPA.
Given the various environmental cost drivers we identified in the case study, and the resulting LCA scores, we have arguably also demonstrated the holistic nature of SOPA: The environmental impact captured here, expressed in a unit-less way, is multi-faceted and not just a set of indicators such as carbon emissions or water use, which are commonly used in other approaches. Thus, evaluation goal 3 is addressed.

\section{Discussion}

\label{sec:discussion}

After presenting SOPA and applying it to a case study, the following section discusses the contribution of this paper and relevant threats to validity. Afterward, the discussion is opened further, and the general context of SOPA is discussed.
 
\subsection{Results}

Since the results of this work are twofold, the case study used for evaluating this contribution will be discussed first, followed by the framework of SOPA itself.

\subsubsection{Case Study}

The application of SOPA to a real-world case as a means of evaluating the contribution and demonstrating its usefulness has indeed shown that SOPA is applicable to assess the environmental impact of business processes and subsequently re-design them. Hence, the practical demonstration addresses \hyperref[RQ3]{RQ3}. Further, by deriving a concrete implementation for the abstract SOPA framework, through which we have conducted the case study, we have addressed \hyperref[RQ2]{RQ2}. However, some threats to validity warrant a discussion: Firstly, the LCA analysis has not been conducted by LCA experts, relies on existing LCA databases and inventory databases, and has been supplemented with reasonable assumptions due to limitations of the conducted interviews. Nonetheless, the quality of the analysis is not highly substantial for the purpose of this case study, which aims at demonstrating the general utility of SOPA --- more exact LCA data would only allow for more precise reasoning about re-design potential, but does not impact the overall analysis procedure. 

Secondly, the evaluation has been done based on simulated event logs, without a real information system behind its execution.
While the use of process simulation instead of real execution data arguably reduces the accuracy of the results, process simulation nonetheless enables the analysis of (potentially fully manual) processes for which no data has been recorded at all, and thus still allows insights to be gained, as has been demonstrated in the case study. Further, we have shown that analyzing and re-designing support processes of organizations, such as the hiring process we used in the evaluation, may indeed contribute to a reduction of the environmental impact of organizations. Concretely, we have described two scenarios that would more than halve the environmental impact of the particular hiring process.
Based on this, we argue that the benefit of analyzing processes for which no execution data exists and deriving meaningful impact reduction re-design scenarios can feasibly be translated to other organizations and other processes.

Moreover, while the use of process expert interviews and the conducted analyses and subsequent simulation have limitations due to their scope, the case study demonstrates that value could be added. While the results should be interpreted carefully, investigating the process even more thoroughly and in conjunction with LCA experts, the confidence in the results can only be increased.
Consequently, the case study shows that, generally, a better elicitation of and reasoning about environmental impact is possible through the application of SOPA to a business process.

\subsubsection{SOPA}
\label{sec:discussion-sopa}

After considering the case study, the following discusses SOPA in general.
Generally, the case study has underlined the capabilities of SOPA to quantify different process re-design measures and to help organizations to identify those which have the greatest impact in reducing environmental costs. Further, it has illustrated how SOPA can enable organizations to model, measure, and optimize their environmental impact in a holistic manner. Thus, with proposing the SOPA framework, we have answered \hyperref[RQ1]{RQ1}.
Nonetheless, some limitations need to be considered here as well.

First, it should be noted that the quality of the results elicited with SOPA depends on the completeness of the input, be it a simulated event log or an extracted one --- the more complete and close-to-real-world the captured behavior and the impact information is, the more accurate the assessment with SOPA is. To increase the exactness of simulation, stochastic information about the investigated process can either be estimated by process experts or identified based on the analysis of event logs~\citep{pufahlDesignExtensibleBPMN2018,wynnBusinessProcessSimulation2008} to increase accuracy. The more in-depth an analyst models the process (e.g., across multiple participants), the more the impact can be attributed to different participants, attributing concrete responsibilities.

Second, the use of LCA scores to compare the environmental impact of activities and processes warrants a discussion. Notably, LCA results can contain a lot of uncertainty, meaning that discrepancies between the actual and the calculated impact may exist --- this needs to be considered during the interpretation phase. The discrepancies could stem, e.g., from erroneous, misspecified, incomplete, or inconsistently elicited data. Relevant literature has identified three main ways of dealing with that uncertainty, which could be further integrated into SOPA: either by improving the quality of the data and models involved, by finding a consensus on data and choices with stakeholders and determining authoritative standards, or by applying statistical methods to incorporate that uncertainty~\citep{finnvedenRecentDevelopmentsLife2009}.

Third, the notion of process re-design that SOPA aims to enable is focused primarily on changing either operational support or improving activity impacts by identifying cost drivers and their impact on the overall process. This does not cover structural re-design in which activities would be replaced or removed, substantially altering the process. However, the authors maintain that a SOPA-led analysis can be combined with other forms of process performance analysis and re-design, allowing further re-design in practice.

Fourth, SOPA does not provide concrete actionable mechanisms to alleviate conflicts between environmental, social, and economic impacts of the analyzed business processes. These may emerge, e.g., when a potential reduction in environmental impact analyzed with SOPA would come at substantial cost to the economic viability of a business, or could harbor undesired social consequences. As argued by \cite{purvisThreePillarsSustainability2019}, there is currently no consensus on how to, formally, characterize sustainability and how its three facets interact. Deciding on how to alleviate conflicts in a business setting (i.e., w.r.t the economic facet of sustainability) requires value judgements \citep{purvisThreePillarsSustainability2019}, which cannot be covered by a technical framework like SOPA. Further, some voices argue for a general incommensurability (i.e., a conceptual difference that makes calculating trade-offs complex or impossible), which would require non-algorithmic multi-criteria frameworks beyond the scope of this work, see \cite{martinez-alierWeakComparabilityValues1998}. %
However, a data-driven approach such as SOPA may allow for a more evidence-based reconciliation of conflicts and assessment of adherence to sustainability-related regulations (e.g.,~\cite{europeancommissionDirective2014952014,europeancommissionRegulation202085}), which may continue to emerge.

Fifth, SOPA differs from other approaches for process optimization (especially algorithmic approaches such as \cite{lowRevisingHistoryCostinformed2016}, \cite{georgoulakosEvolutionaryMultiObjectiveOptimization2017} and \cite{vergidisBusinessProcessDesign2008}), where optimal process designs or alternative execution scenarios are identified based on i.a. costs of cases, activities, resources, the input-output relation of activities and resources, and the runtime of activities.
There, 1) commensurability between various costs through a common cost unit (such as monetary costs in terms of US-\$) is assumed, and 2) a uni-directional relation between process parameters and costs is assumed --- changing process parameters leads to a change in costs, which are to be optimized by way of changing the parameters. SOPA however aims at measuring the environmental impact holistically, so that, as discussed above, a comparison with e.g., other cost units is potentially infeasible, and assumes a bidirectional relation between concrete cost drivers of activity instances and process behavior, as illustrated in the third facet of process re-design in Section~\ref{sec:results-frame-re}. More concretely, the way in which cost drivers can be changed is heavily constrained by expert knowledge and dependent upon LCA analyses. Only after identifying potential for improvement, the process experts can identify other concretizations of abstract cost drivers that could lead to a reduction in impact and calculate their LCA values. This is due to the fact that there might exist an interaction between specific environmental cost drivers and the process behavior. This means that they cannot be arbitrarily combined during actual process execution, but that instead expert knowledge is needed to determine whether changes can or cannot be implemented, and to what effect w.r.t. the behavior of the process. Further, changing the probability of cost variants represents changing the implementation of the business process, and therefore the feasibility of such an optimization needs to be considered by the process expert. 
Consequently, reducing the environmental impact itself is not just an algorithmic problem to which an optimal solution w.r.t. the process domain can be derived, but one that also requires an in-depth understanding and deliberation of the domain. %

And finally, when considering other methods that aim at capturing impacts on environmental sustainability (e.g.,~\cite{elkingtonACCOUNTINGTRIPLEBOTTOM1998,ellramFrameworkTotalCost1993}), the advantages and limitations of SOPA need to be stressed: The scope of SOPA is bound by the scope of LCA -- other approaches focus only on individual enterprises or processes and are, as opposed to SOPA and LCA, not intended to cross organizational boundaries. Further, other approaches are focused on particular indicators that are, as discussed before, unable to capture the entire environmental impact of processes or organizations, as opposed to LCA. Notably, the inherent capabilities for sustainability-oriented process re-design by leveraging process simulation is a differentiator to other approaches, which assume a static perspective, and do not provide explicit mechanisms for re-design.

\subsection{General Discussion}

Subsequently, after discussing SOPA, this section further opens the discussion and considers the context of this work.

\subsubsection{Sustainability through Optimization?}

Firstly, while this work itself offers a tool for optimizing (i.e., by reducing) the environmental impact of business processes, the idea behind process optimization could be considered critically. When optimizing towards a specific goal, in this case the reduction of environmental impact, the measure with which this optimization is tracked should be discussed --- any optimization should always serve towards the end goal of impact reduction, and not just an optimization for the metric's sake~\citep{vanderkolkNumbersSpeakThemselves2022}. The actual prevention therefore depends on the context of application of SOPA, not just the application of SOPA itself, in other words, in which manner SOPA is applied in practice: is it just applied to reduce the impact scores as much as possible, or is it applied to actually \textit{understand} the environmental impact and reduce it accordingly? This closely relates to rebound effects (see~\cite{herringEnergyEfficiencyCritical2006}), where, through greater efficiency of e.g., a service, greater use is encouraged, and consequently practically all benefits are absorbed. Also of relevance is the discussion of energy efficiency provided by~\cite{shoveWhatWrongEnergy2018}, where efficiency measures need to be critically examined for what they measure, and how they perpetuate the status quo, leading to path dependencies and preventing long-term reduction of environmental impact.

\subsubsection{Sustainability through BPM?}

Moreover, the authors of this work would like to open a discussion on the extent with which the tools and techniques of BPM are even applicable for improving the sustainability of business processes.

As argued by \cite{gopelGreatMindshift2016}, an increasing integration of environmental, social and economic concerns led to the dispersion of an \textit{``economic mind-set''} into the governance of an increasing number of areas of life, thereby solidifying path dependencies towards unsustainability. Therefore, a critical view might be necessary: How much does the application of BPM in order to address problems of environmental sustainability contribute to that solidification? This warrants further research and discussion. Other areas, such as sustainability transitions research (STR)~\citep{feolaCapitalismSustainabilityTransitions2020} move towards critically examining the role of \textit{capitalism} (i.e., a specific form of social and economic organization going beyond the economic sphere~\citep{feolaCapitalismSustainabilityTransitions2020}) so that truly sustainable transitions can be enabled.
As a recent study about a large oil company has shown, accurate scientific insights into climate change (e.g., accurate predictions of carbon budget) were systematically withheld and miscommunicated to the outside~\citep{supranAssessingExxonMobilGlobal2023}, thus hinting at how economic incentives might actively hinder %
sustainability transformations.

Following \cite{gopelGreatMindshift2016}, the governance of nature would require, instead of the prevalent ``monetized cost-benefit analyses'', multidimensional evidence. Unlocking path dependencies towards sustainability arguably is hindered by inequalities inherent to ``market logic, laws, and institutions [that are being] created, overlooked, or disguised [through] undifferentiated cost–benefit and growth analyses''~\citep{gopelGreatMindshift2016}. Further, sustainability solutions may not lie in increasing the efficiency of technologies and the effectiveness of economic incentives, but instead may need to be facilitated by organizational setups and sociocultural frameworks defining the purpose of technologies and economic instruments~\citep{gopelGreatMindshift2016}. Thus, while SOPA is a contribution within BPM to enable further considerations of sustainability with it, the overall role of BPM itself might need to be investigated closely for how it has internalized mechanisms that might hinder adequately addressing unsustainable path dependencies, and how it might be leveraged for sustainability solutions within institutions.

One interesting avenue of research might be applying the research paradigm of \textit{critical research} (i.e., a paradigm aiming at critically examining a subject and its context, and investigating its hidden limitations to change, imposed by economic, political, and cultural authority~\citep{howcroftHandbookCriticalInformation2005}) to BPM, an approach that has found some application in IS research.
Moreover, the concept of \textit{sufficiency}, focusing on \textit{deceleration} (less speed and efficiency), \textit{deglobalization} (less fragmented cross-global productions chains that externalize environmental impacts), \textit{decluttering}, and \textit{decommodification} (moving away from commodified mental models focused on numbers and financial benefits)~\citep{schneidewindInstitutionalFrameworkSufficiency2014,gopelGreatMindshift2016}, as alternatives to existing economic objectives, might be able to contribute towards a more sustainability-oriented derivative of BPM.

Finally, it should be noted that, as the discussion above has made clear, thorough considerations and even transformations of BPM and the context in which it is applied are necessary for facing the grand challenge of addressing sustainability --- but they can only be achieved step by step. The authors would like to underline that the contribution of SOPA could potentially constitute such a step.

\subsection{Future Work}

Finally, this section considers future work and research opportunities highlighted by this contribution. 
One interesting line of research would be, how (un)certainty and confidence intervals could be included in the analysis and simulation, to do the uncertainty inherent to LCA and process simulation justice. This could indicate whether ``the general direction of improvement'' is correct or not, instead of implying absolute certainty. Further, this resembles the reality of process execution more closely, where variations in impact exist due to minute fluctuations and variations in process execution, while still providing a benefit for impact analysis (e.g., one process variant is 30\% to 33\% less impactful than the other).

Another fruitful avenue of research might lie in comparing SOPA with other frameworks that assess the environmental impact of business processes regarding more user-focussed aspects, such as user satisfaction or comprehensibility of the results. Since this work's main purpose is to provide a formal point of reference for implementing SOPA in practice, we believe that such practical considerations require a more rounded implementation than the prototype provided herein. Therefore, we refer this investigation to future work.

Moreover, a more systematic tracking of sustainability factors can ameliorate both analysis results and simulation models: a ``data mining for LCA'' (e.g., based on sensor data) could form a data-driven basis for the analysis, by either supplementing existing LCA information with sensor data recorded in IS, or by mining relevant data from process executions to make LCA analysis more exact. Further, the simulation could be made more precise by extracting accurate simulation parameters based on observed behavior. A need for these data-driven approaches is only underlined by the discussion about the actual efficacy of carbon offset certificates\footnote{See \url{https://www.theguardian.com/environment/2023/jan/18/revealed-forest-carbon-offsets-biggest-provider-worthless-verra-aoe} and \url{https://www.source-material.org/vercompanies-carbon-offsetting-claims-inflated-methodologies-flawed/} [Accessed: July 17, 2024]}, where a surplus of emissions is still caused, and the actual offset is not closely in relation to what is supposed to be offset, due to limited traceability~\citep{westActionNeededMake2023}.

Furthermore, the exactness of LCA could be increased by encouraging cooperation of manufactures and produces within and across industries to enable sharing of more exact LCA values. While some industry-specific databases have been emerging (see~\cite{finnvedenRecentDevelopmentsLife2009}), further proliferation can only be helpful. Moreover, following the strength of LCA to assess impacts across organizational boundaries, investigating how this approach could be extended to process collaborations or choreographies could be fruitful. Notably, \cite{thiesPotentialNetworkCentricSolution2012} outlines a network-centric sharing system for environmental performance indicators across enterprises, which could be extended with LCA databases and results of SOPA analyses.

With respect to the other two dimensions of sustainability not yet addressed by SOPA, a combination with other frameworks for process analysis and optimization might be feasible, allowing for optimization towards, e.g., ecological \textit{and} economic facets. Concretely, literature has proposed the \textit{Life Cycle Sustainability Assessment} (LCSA, see \cite{finkbeinerLifeCycleSustainability2010}) and \textit{Holistic and Integrated LCSA} (HILCSA, see \cite{zeugLifeCycleSustainability2023}) approaches. Both are extensions of traditional LCA that aim at incorporating all three facets of sustainability into one assessment. As to the discussion of Section~\ref{sec:discussion-sopa}, LCSA and HILCSA also require a weighting of the three dimensions of sustainability, and thus, a value judgement. Integrating these approaches with SOPA by including an explicit non-technical mechanism for performing such a weighting could potentially allow for considering all three facets of sustainability at once during process optimization, and could be a promising avenue of research.

Additionally,~\cite{naumannGREENSOFTModelReference2011} propose a model with which sustainability concerns are explicitly extended to software products so that software products can be developed, operated and used in a sustainable manner. This practice, also known as \textit{green coding}, could be employed to further the scope of analysis around the analyzed business processes, by e.g., improving process engines or other IT systems employed in process execution, especially when they contribute heavily to execution of individual activities or processes. %

Further, the emerging technique of object-centric event logs~\citep{ghahfarokhiOCELStandardObjectCentric2021}, where instead of a flat stream of events belonging to individual cases, events across several objects are recorded without a singular case notion during process execution, appears promising: that way, the impact of objects across their life cycle and their contribution during process execution might be captured even more precisely, closely relating to the life cycle aspect of LCA. This, however, would require extended tool- and simulation support, given the young nature of this technique.

Finally, we would like to underline that the area of BPS is still an area of ongoing research, where continuous improvements are being made --- future improvements, so that process behavior can be simulated even more accurately, can only be a positive contribution to SOPA. Clearly, improving simulation leads to better results and analysis of the environmental impact of business processes.

\section{Conclusion}

Given the accelerating human-made climate change, businesses have an increasing need to assess, monitor, and improve the environmental impact of their business processes. To this end, we have combined Life Cycle Assessment and Activity-based Costing for Business Process Management purposes with a Design Science Research-based approach. The resulting framework of SOPA fills a gap in existing works, which we have found to be focussed on a limited set of environmental indicators and limited in their capability to be used for sustainability-oriented process re-design. We have applied SOPA to a case study, and shown its potential, in conjunction with Business Process Simulation, to contribute towards monitoring and re-designing a business process to be less environmentally impactful. Moreover, we have discussed how SOPA might benefit from additional environmental data and more accurate simulation models. While SOPA provides a technical solution for environmental impact analysis and re-design of business processes, the overall role of BPM in contributing towards environmentally unsustainable business activities also needs to be considered. SOPA is a way of integrating environmental considerations into BPM, but its underlying mechanisms that might prevent even more sustainable transformations warrant a critical investigation, in order to transition towards a more sustainable future.

\backmatter

\begin{appendices}

\section{Illustrating Example}\label{sec:appendix0}

In the following, we provide further examples of the concepts we formalized in \autoref{sec:results-frame}. We stress that these examples primarily serve to illustrate the concepts, and are explicitly \textit{not} meant for completeness -- for this, we refer to the case study in \autoref{sec:results-eval} and the corresponding supplementary material available online.\footnote{\url{https://doi.org/10.6084/M9.FIGSHARE.22591513} [Accessed: July 17, 2024]}

\subsection{Environmental Cost Driver Hierarchy and Cost Function}
\label{sec:example-hierarchy-cost}

Recall that Definition \ref{def:hierarchy} describes into which concrete cost drivers abstract ones can be concretized during process execution, and that Definition \ref{def:cost-fn} assigns a cost, representing the LCA-derived environmental impact score, to each concrete cost driver.
The abstract cost driver $ d\in D$, for example the delivery of in-house mail, could have various concretizations defined in the cost driver hierarchy $H$, such as $c_{1}, c_{2}\in C$. Both represent different concretizations of the abstract delivery of in-house mail. To show that both concrete cost drivers concretize the abstract one, in other words, that $(d,c_1),(d,c_2) \in H$, we also write $c_{1d}, c_{2d}$. In particular, $c_{1d}$ might be the delivery over a certain distance via a specific car, whereas $c_{2d}$ might be the delivery over the same distance with a specific electric bicycle. Their difference in environmental impact is reflected by the $cost$ function, which, in this example, might assign a cost of $3.91\times10^{-5}$ to the first and $4.22\times10^{-6}$ to the second concrete cost driver.

\subsection{Cost Variant Configuration}
\label{sec:example-cost-variant}

The following example Listing \ref{lst:cost-variant-excerpt} illustrates how, based on our prototypical implementation of SOPA, the concretization of cost drivers in process instances during process simulation may be configured. Each variant describes how a set of abstract cost drivers will be concretized so that upon assignment to an activity instance, a concrete impact score can be determined for the driver, and subsequently, the activity instance. Note that the \textit{actual} abstract cost drivers are provided in an annotated process model, see e.g.\ Fig.~\ref{fig:hiring_process_driver}.

\begin{lstlisting}[language=XML, frame=tb, basicstyle=\linespread{1}\footnotesize, tabsize=2, breaklines=true, showtabs=false, showstringspaces=false, caption=Excerpt of a cost variant configuration file describing how abstract environmental cost drivers are concretized depending on the respective cost variant of a process instance., numbers=none,label=lst:cost-variant-excerpt,captionpos=t]
<?xml version="1.0" encoding="UTF-8"?>
<costVariantConfig count="500">
    <variant id="standard procedure" frequency="0.5">
        <driver id="Request for job advertisement" cost="0.0000289"/>
        <driver id="In-house mail" cost="0.0000391"/>
        <driver id="Advertisement" cost="0.0000291"/>
        ...
    </variant>
    <variant id="transport with e-bike" frequency="0.2">
        <driver id="Request for job advertisement" cost="0.0000289"/>
        <driver id="In-house mail" cost="0.00000422"/>
        <driver id="Advertisement" cost="0.0000291"/>
        ...
    </variant>
    <variant id="digital only" frequency="0.3">
        <driver id="Request for job advertisement" cost="0.0000195"/>
        <driver id="In-house mail" cost="0.000000151"/>
        <driver id="Advertisement" cost="0.0000195"/>
        ...
  </variant>
</costVariantConfig>
\end{lstlisting}

Concretely, the first variant shown in Listing \ref{lst:cost-variant-excerpt}, representing the \textit{standard procedure} of the running example, will be simulated 250 times out of the configured 500 process instances. The environmental cost drivers all provide \textit{concrete} impact scores, and will be assigned to all relevant activity instances that are simulated with the first cost variant, based on the cost driver hierarchy and annotated process model. The second cost variant, representing in this example a different mode of transport being used, will be simulated 100 times. Notably, the concrete environmental cost driver \textit{In-house mail} differs from the standard procedure, illustrating a different concretization in the respective process instances. All other concrete environmental cost drivers stay the same, which can be seen by the ``cost'' field. Finally, the digital-only variant completely differs in the concrete scores of the concrete environmental cost drivers, representing process instances where only digital documents are used, which can also be seen in the cost of all environmental cost drivers being different.

\subsection{Event Log Excerpt}
\label{sec:example-event-log}

Listing~\ref{lst:xes-excerpt} displays an \textit{example} event log excerpt for the hiring process, which was created with our prototypical implementation for the case study evaluation. Notably, Definition \ref{def:process-instance} requires a total ordering of activities. While the timestamps in Listing~\ref{lst:xes-excerpt} would only suffice for partial ordering, the XES standard \citep{IEEEStandardEXtensible} describes that recorded activity instances of a process instance are, in fact, ordered. In practice, a pre-determined ordering for activity instances with the same timestamp within one process instance is usually assumed, maintaining the total order requirement \citep{vdaalstConcurrency2021,rebmanUncovering2024}.

\begin{lstlisting}[language=XML, frame=tb, basicstyle=\linespread{1}\footnotesize, tabsize=2, breaklines=true, showtabs=false, showstringspaces=false, caption=Excerpt of an XES~\citep{IEEEStandardEXtensible} event log belonging to the hiring process showing additional cost variant and environmental cost driver information., numbers=none,label=lst:xes-excerpt,captionpos=t]
<trace>
		<string key="concept:name" value="410"/>
		<string key="cost:variant" value="standard procedure"/>
		<event>
			<string key="concept:name" value="Hiring required"/>
			<string key="lifecycle:transition" value="start"/>
			<date key="time:timestamp" value="2026-07-17T15:35:28+02:00"/>
		</event>
		<event>
			<string key="concept:name" value="Hiring required"/>
			<string key="lifecycle:transition" value="complete"/>
			<date key="time:timestamp" value="2026-07-17T15:35:28+02:00"/>
		</event>
		<event>
			<string key="concept:name" value="Submit request for job advertisement (Department)"/>
			<string key="lifecycle:transition" value="start"/>
			<date key="time:timestamp" value="2026-07-17T15:35:28+02:00"/>
		</event>
		<event>
			<string key="cost:driver" value="Request for job advertisement"/>
			<string key="concept:name" value="Submit request for job advertisement (Department)"/>
			<string key="lifecycle:transition" value="complete"/>
			<date key="time:timestamp" value="2026-07-17T16:12:16+02:00"/>
		</event>
    	...
</trace>
\end{lstlisting}

\subsection{Activity Instances}
\label{sec:example-activity-instance}

Per Definition \ref{def:activity-instance}, each activity instance consists of an activity and a set of concrete environmental cost drivers. This means that during execution or simulation, each activity execution is recorded with a set of concrete cost drivers that represent what concretely contributed to the environmental impact of that one activity execution.

For example, the environmental cost of an activity $a \in A$ (such as an activity in which a set of documents is received and checked) could be represented by the abstract cost driver $d \in D$, symbolizing for instance the delivery of a document via the in-house mail. If the cost driver hierarchy $H$ described that $d$ could either be concretized into $c_{1_d}$ or $c_{2_d}$ $\in C$, instances of activity $a$ would either occur with $c_{1_d}$ as $(a,\{c_{1_d}\})$, or $(a,\{c_{2_d}\})$, representing the execution of activity $a$ at runtime that incurs different costs based on the involved concrete cost drivers $c_{1_d}$ and $c_{2_d}$. For instance, $c_{1_d}$ could represent the delivery of a set of documents via a petrol-powered car over a distance of 750m, while $c_{2_d}$ could represent the delivery of the same set of documents over the same distance with an electric bike. This, notably, also works for sets of abstract respective concrete cost drivers associated with activities.

\subsection{Process Instances}
\label{sec:example-process-instance}

Following Definition \ref{def:process-instance}, process instances consist of sequences of activity instances, similar to the example above. Assume a process consisting of two activities $a, b \in A$, where for example activity $a$ might deal with the selection of candidates, and activity $b$ with interviewing a candidate in a hiring process. Further, assume that activity $a$ is associated with abstract cost drivers $d_1, d_2 \in D$, and activity $b$ with $d_3 \in D$. For example, $d_1$ might represent the cost of sifting candidates, whereas $d_2$ might represent the cost of compiling a list of candidates, and $d_3$ the cost of conducting a single interview. Further, assume that the abstract cost drivers $d_1, d_2, d_3$ can be concretized into concrete cost drivers $c_{1d_1}, c_{2d_1}, c_{3d_2}, c_{4d_2}, c_{5d_2}, c_{6d_3}, c_{7d_2} \in C$. As in \autoref{sec:example-hierarchy-cost}, we indicate in the indices of the concrete cost drivers the abstract ones they are associated to in the cost driver hierarchy $H$. For example, two different concretizations of $d_1$ may represent two different concrete ways of sifting the candidates (either paper-based or on a computer), and consequently, whereas $d_2$ may be concretized into three different concrete modes of compiling a list of candidates, with different environmental impacts. Further, the interview may be conducted either on-site or online, leading to different concretizations of $d_3$ with their own impact score. Consequently, $t_1 = \langle(a,\{c_{1d_1},c_{5d_2}\}),(b,\{c_{6d_3}\})\rangle$ and $t_2 = \langle(a,\{c_{2d_1},c_{3d_2}\}),(b,\{c_{6d_3}\})\rangle$  with $t_1, t_2 \in I^*$ would be two process instances that represent the concrete environmental impact incurred at runtime by the various activity instances and hence, the process instances.

\subsection{Environmental Activity Instance Cost}
\label{sec:example-activity-instance-cost}

In line with Definition \ref{def:activity-instance-cost}, the cost in terms of environmental impact of an activity instance can be calculated by considering the cost of the concrete cost drivers associated with that activity instance.
More concretely, consider an activity instance $(a,\{c_1,c_2\})$, with $a \in A$, $c_1, c_2 \in C$, similar to \autoref{sec:example-process-instance}, of activity $a$, which might represent the selection of candidates in a hiring process. The $cost$ function assigns a concrete score to each concrete cost driver, and can be used to calculate the overall cost of the activity instance. If $cost(c_1)=u$, and $cost(c_2)=v$, with $u,v \in \mathbb{Q}$, the overall environmental cost of that activity instance $\mi{activity\_instance\_cost}$ is equal to the sum of all costs of the involved concrete cost drivers, being $u+v$. More concretely, assume that the concrete cost driver $c_1$, representing a very concrete way of sifting the candidates using a computer, has a $cost$ value of $5.85\times10^{-5}$, and the concrete cost driver $c_2$, representing a very concrete way of compiling a final list of candidates, has a $cost$ value of $1.34\times10^{-5}$. Then, the overall environmental impact score $\mi{activity\_instance\_cost}(a,\{c_1,c_2\})$ of the one instance of activity $a$ would be equal to $7.19\times10^{-5}$.

\subsection{Environmental Process Instance Cost}
\label{sec:example-process-instance-cost}

As per Definition \ref{def:process-instance-cost}, we can calculate the environmental impact of a process instance based on the environmental impact of all activity instances belonging to that process instance. Consider, for example, a process instance $t_1 = \langle(a,\{c_{1d_1},c_{5d_2}\}),(b,\{c_{6d_3}\})\rangle \in I^*$ from \autoref{sec:example-process-instance}. Calculating the $\mi{activity\_instance\_cost}$ of each activity instance of $t_1$ and subsequently the sum of those yields the $\mi{process\_instance\_cost}$, representing the environmental impact score of that one process instance. More concretely, if $\mi{activity\_instance\_cost}(a,\{c_{1d_1},c_{5d_2}\}) = 7.19\times10^{-5}$ and $\mi{activity\_instance\_cost}(b,\{c_{6d_3}\}) = 3.5\times10^{-5}$, the overall $\mi{process\_instance\_cost}(t_1)$ is equal to $10.69\times10^{-5}$.

\subsection{Average Environmental Activity Instance Cost}
\label{sec:example-average-activity-instance-cost}
Definition \ref{def:average-activity-cost} describes how, for a specific activity of a process, the average impact of that activity across all activity executions can be calculated. 
Described more concretely, by counting the activity instances of an activity $a \in A$ with all combinations of possible concrete cost drivers, multiplying this with the respective $\mi{activity\_instance\_cost}$ and dividing by the total number of occurrences of $a$, we can calculate the average cost of activity $a$ across all instances, taking into account all possible concretizations of the abstract cost drivers that have occurred. For example, consider three process instances $t_1 = \langle(a,\{c_{1d_1},c_{5d_2}\}),(b,\{c_{6d_3}\})\rangle$, $t_2 = \langle(a,\{c_{2d_1},c_{3d_2}\}),(b,\{c_{6d_3}\}),(a,\{c_{2d_1},c_{3d_2}\})\rangle, t_3 = \langle(a,\{c_{2d_1},c_{3d_2}\}),(b,\{c_{6d_3}\})\rangle$  with $t_1, t_2, t_3 \in I^*$. Now, the \textit{average} environmental activity instance cost of $a$ can be calculated via the environmental activity instance costs of $a$, being $\mi{activity\_instance\_cost}(a,\{c_{1d_1},c_{5d_2}\})$ and $\mi{activity\_instance\_cost}(a,\{c_{2d_1},c_{3d_2}\})$. For all possible concretizations of cost drivers in $H$ (here, i.a. $\{c_{1d_1},c_{5d_2}\}$ and $\{c_{2d_1},c_{3d_2}\}$), we multiply how many times we observe that concretization for activity $a$ in a particular process instance (thereby taking looping behavior into account) with the $\mi{activity\_instance\_cost}$ of $a$ and that concretization. This is divided by the overall number of instances of $a$, yielding the environmental activity cost on average across all process instances. In this particular case, assuming an $\mi{activity\_instance\_cost}(a,\{c_{1d_1},c_{5d_2}\}) = 7.19\times10^{-5}$ and $\mi{activity\_instance\_cost}(a,\{c_{2d_1},c_{3d_2}\}) = 3.63\times10^{-5}$, we calculate the average environmental activity instance cost as $\mi{average\_activity\_cost}(a, \{t_1,t_2,t_3\}) = \frac{1\cdot7.19\times10^{-5}+2\cdot3.63\times10^{-5}+1\cdot3.63\times10^{-5}}{1 + 2+ 1} = 4.52\times10^{-5}\times$. For clarity, we leave out all those possible concretizations that do not appear in the activity instances of $a$ and would have a $\mi{specific\_count}$ of 0.

\subsection{Average Environmental Process Instance Cost}
\label{sec:example-average-process-instance-cost}
Similar to the example above, Definition \ref{def:average-process-instance-cost} lays out how the average environmental cost of process instances can be calculated. Assume three process instances $t_1, t_2, t_3 \in I^*$, and $T = \{t_1, t_2, t_3\}$. Further, assuming that the environmental process instance cost per Definition \ref{def:process-instance-cost} for these instances is $\mi{process\_instance\_cost}(t_1) = 10.69\times10^{-5}$, $\mi{process\_instance\_cost}(t_2) = 13.31\times10^{-5}$, and $\mi{process\_instance\_cost}(t_3) = 9.00\times10^{-5}$. Then, $\mi{average\_process\_instance\_cost}(T) = \frac{10.69\times10^{-5}+ 13.31\times10^{-5} + 9.00\times10^{-5}}{3} = 11\times10^{-5}$.

\section{Additional Figures}
\label{sec:appendix1}

The following provides additional details of the data elicited during the evaluation of SOPA.

\begin{table}[ht]
\caption{Environmental activity and process instance costs of the different scenarios evaluated with SOPA in \autoref{sec:results-eval}\label{tab:sim-results}}
    \begin{tabular*}{\textwidth}{c|ccc}
\toprule
\textbf{Parameter}	& \textbf{Scenario A} & \textbf{Scenario B} & \textbf{Scenario C}\\
\midrule
Check contents of advertisement req. (SC) & $3.91\times10^{-5}$& $4.22\times10^{-6}$& $1.51\times10^{-7}$\\\midrule
Check contents of hiring req. (SC) & $3.91\times10^{-5}$& $4.22\times10^{-6}$& $1.51\times10^{-7}$\\\midrule
Formally assess hiring req. (HR) & $3.91\times10^{-5}$& $4.22\times10^{-6}$& $1.51\times10^{-7}$\\\midrule
Formally assess hiring req. (Faculty) & $3.91\times10^{-5}$& $4.22\times10^{-6}$& $1.51\times10^{-7}$\\\midrule
Finalize contract (HR) & $2.54\times10^{-5}$& $2.54\times10^{-5}$& $1.95\times10^{-5}$\\\midrule
Submit request for job advertisement (Dep) & $2.89\times10^{-5}$& $2.89\times10^{-5}$& $1.95\times10^{-5}$\\\midrule
Check contents of advertisement req. (DO)  & $3.91\times10^{-5}$& $4.22\times10^{-6}$& $1.51\times10^{-7}$\\\midrule
Sift and select candidates (Dep) & $5.85\times10^{-5}$& $5.85\times10^{-5}$& $2.93\times10^{-5}$\\\midrule
Check contents of advertisement req. (WR)  & $3.91\times10^{-5}$& $4.22\times10^{-6}$& $1.51\times10^{-7}$\\\midrule
Submit job advertisement (HR) & $2.91\times10^{-5}$& $2.91\times10^{-5}$& $1.95\times10^{-5}$\\\midrule
Check contents of hiring req. (DO) & $3.91\times10^{-5}$& $4.22\times10^{-6}$& $1.51\times10^{-7}$\\\midrule
Formally assess advertisement req. (Faculty) & $3.91\times10^{-5}$& $4.22\times10^{-6}$& $1.51\times10^{-7}$\\\midrule
Formally assess advertisement req. (HR) & $3.91\times10^{-5}$& $4.22\times10^{-6}$& $1.51\times10^{-7}$\\\midrule
Check contents of hiring req. (WR) & $3.91\times10^{-5}$& $4.22\times10^{-6}$& $1.51\times10^{-7}$\\\midrule
Request hiring of candidate (Dep) & $4.31\times10^{-5}$& $4.31\times10^{-5}$& $1.95\times10^{-5}$\\\midrule
Conduct interview with candidate & $3.50\times10^{-5}$& $3.50\times10^{-5}$& $3.50\times10^{-5}$\\\midrule\midrule
Average Environmental Process Instance Cost & $7.00\times10^{-4}$& $3.48\times10^{-4}$&$2.44\times10^{-4}$\\
\bottomrule
\end{tabular*}
\end{table}

\begin{table}[!ht]
\caption{Relative differences in environmental activity and process instance costs of the different scenarios in \autoref{sec:results-eval}\label{tab:sim-results-compare}}
\begin{tabular*}{\textwidth}{c|ccc}
\toprule
\textbf{Parameter}	& \textbf{Scenario A --- B} & \textbf{Scenario A --- C} \\
\midrule
Check contents of advertisement req. (SC) & $-89.2\% $ & $-99.61\% $ \\\midrule
Check contents of hiring req. (SC) & $-89.2\% $ & $-99.61\% $ \\\midrule
Formally assess hiring req. (HR) & $-89.2\% $ & $-99.61\% $ \\\midrule
Formally assess hiring req. (Faculty) & $-89.2\% $ & $-99.61\% $ \\\midrule
Finalize contract (HR) & $0\% $ & $-23.22\% $ \\\midrule
Submit request for job advertisement (Dep) & $0\% $ & $-32.52\% $ \\\midrule
Check contents of advertisement req. (DO)  & $-89.2\% $ & $-99.61\% $ \\\midrule
Sift and select candidates (Dep) & $0\% $ & $-50.00\%$ \\\midrule
Check contents of advertisement req. (WR)  & $-89.2\% $ & $-99.61\% $ \\\midrule
Submit job advertisement (HR) & $0\% $ & $-32.98\% $ \\\midrule
Check contents of hiring req. (DO) & $-89.2\% $ & $-99.61\% $ \\\midrule
Formally assess advertisement req. (Faculty) & $-89.2\% $ & $-99.61\% $ \\\midrule
Formally assess advertisement req. (HR) & $-89.2\% $ & $-99.61\% $ \\\midrule
Check contents of hiring req. (WR) & $-89.2\% $ & $-99.61\% $ \\\midrule
Request hiring of candidate (Dep) & $0\% $ & $-54.75\% $ \\\midrule
Conduct interview with candidate & $0\% $ & $0\% $ \\\midrule\midrule
Average Environmental Process Instance Cost & $-50.18\%$ & $-65.08\%$ \\
\bottomrule
\end{tabular*}
\end{table}

\clearpage

\begin{figure}[htbp]
    \centering
    \includegraphics[width=0.9\textwidth]{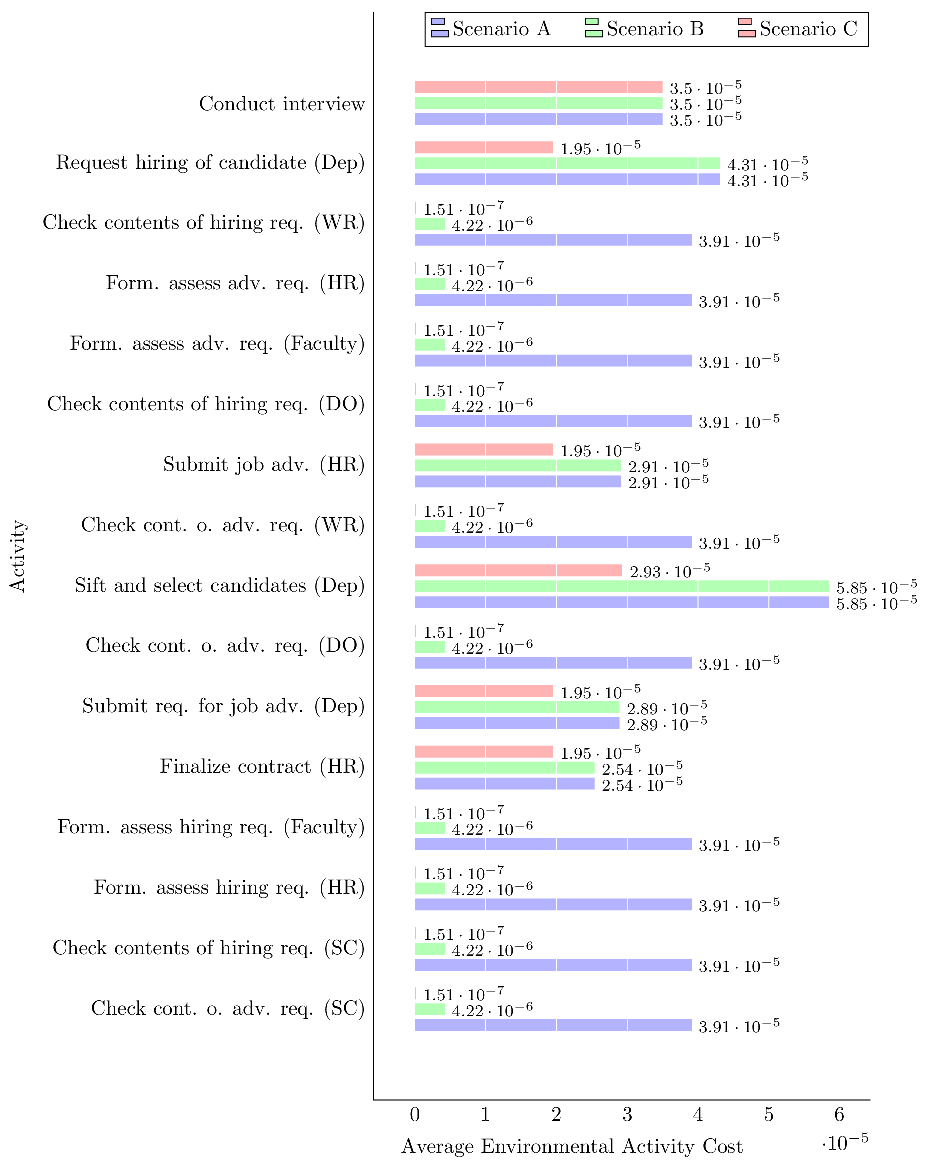}
    \caption{Bar chart displaying the average environmental activity costs per activity and per scenario resulting from evaluating SOPA with the case study, see \autoref{sec:results-eval}}
    \label{fig:bar_results}
\end{figure}

\clearpage

\end{appendices}

\bibliography{sn-article}%

\begin{thebibliography}{91}
\providecommand{\natexlab}[1]{#1}
\providecommand{\url}[1]{{#1}}
\providecommand{\urlprefix}{URL }
\providecommand{\doi}[1]{\url{https://doi.org/#1}}
\providecommand{\eprint}[2][]{\url{#2}}
 \bibcommenthead

\bibitem[{Aguirre et~al.(2013)Aguirre, Parra, and
  Alvarado}]{aguirreCombinationProcessMining2013}
Aguirre S, Parra C, Alvarado J (2013) Combination of {{Process Mining}} and
  {{Simulation Techniques}} for {{Business Process Redesign}}: {{A
  Methodological Approach}}. In: {van der Aalst} WMP, Mylopoulos J, Rosemann M,
  et~al. (eds) Data-{{Driven Process Discovery}} and {{Analysis}}, vol 162.
  {Springer}, {Berlin, Heidelberg}, p 24--43, \doi{10.1007/978-3-642-40919-6_2}

\bibitem[{Bauer(2008)}]{bauerLeitbildNachhaltigenEntwicklung2008}
Bauer S (2008) Leitbild der {{Nachhaltigen Entwicklung}}. In: Informationen zur
  politischen Bildung. Bundeszentrale f\"ur politische Bildung, Bonn

\bibitem[{Borkowski et~al.(2019)Borkowski, Fdhila, Nardelli, Rinderle-Ma, and
  Schulte}]{borkowskiFailure2019}
Borkowski M, Fdhila W, Nardelli M, et~al. (2019) Event-based failure prediction
  in distributed business processes. Information Systems 81:220--235.
  \doi{https://doi.org/10.1016/j.is.2017.12.005}

\bibitem[{vom Brocke et~al.(2012)vom Brocke, Seidel, and
  Recker}]{vombrockeGreenBusinessProcess2012}
vom Brocke J, Seidel S, Recker J (eds)  (2012) Green {{Business Process
  Management}}: {{Towards}} the {{Sustainable Enterprise}}. {Springer},
  {Berlin, Heidelberg}, \doi{10.1007/978-3-642-27488-6}

\bibitem[{Brooks et~al.(2012)Brooks, Wang, and
  Sarker}]{brooksUnpackingGreenReview2012}
Brooks S, Wang X, Sarker S (2012) Unpacking {{Green IS}}: {{A Review}} of the
  {{Existing Literature}} and {{Directions}} for the {{Future}}. In: vom Brocke
  J, Seidel S, Recker J (eds) Green {{Business Process Management}}.
  {Springer}, {Berlin, Heidelberg}, p 15--37, \doi{10.1007/978-3-642-27488-6_2}

\bibitem[{Brundtland(1987)}]{brundtlandOurCommonFuture1987}
Brundtland GH (1987) Our {{Common Future}}\textemdash{{Call}} for {{Action}}.
  Environmental Conservation 14(4):291--294. \doi{10.1017/S0376892900016805}

\bibitem[{Cleven et~al.(2012)Cleven, Winter, and
  Wortmann}]{clevenManagingProcessPerformance2012}
Cleven A, Winter R, Wortmann F (2012) Managing {{Process Performance}} to
  {{Enable Corporate Sustainability}}: {{A Capability Maturity Model}}. In: vom
  Brocke J, Seidel S, Recker J (eds) Green {{Business Process Management}}.
  {Springer}, {Berlin, Heidelberg}, p 111--129,
  \doi{10.1007/978-3-642-27488-6_7}

\bibitem[{Cooper and Kaplan(1991)}]{cooperDesignCostManagement1991}
Cooper R, Kaplan RS (1991) The Design of Cost Management Systems: Text, Cases,
  and Readings. {Prentice Hall}

\bibitem[{Couckuyt and Van~Looy(2019)}]{couckuytGreenBPMBusinessOriented2019}
Couckuyt D, Van~Looy A (2019) Green {{BPM}} as a {{Business-Oriented
  Discipline}}: {{A Systematic Mapping Study}} and {{Research Agenda}}.
  Sustainability 11(15):4200. \doi{10.3390/su11154200}

\bibitem[{{de Bruijn} et~al.(2002){de Bruijn}, {van Duin}, Huijbregts, Guinee,
  Gorree, Heijungs, Huppes, Kleijn, {de Koning}, {van Oers}, Wegener~Sleeswijk,
  Suh, and {Udo de Haes}}]{debruijnHandbookLifeCycle2002}
{de Bruijn} H, {van Duin} R, Huijbregts MAJ, et~al. (eds)  (2002) Handbook on
  {{Life Cycle Assessment}}: {{Operational Guide}} to the {{ISO Standards}},
  Eco-{{Efficiency}} in {{Industry}} and {{Science}}, vol~7. {Springer
  Netherlands}, {Dordrecht}, \doi{10.1007/0-306-48055-7}

\bibitem[{D{\'i}az et~al.(2019)D{\'i}az, Settele, Brond{\'i}zio, Ngo,
  Gu{\`e}ze, Agard, Arneth, Balvanera, Brauman, Butchart, Chan, Garibaldi,
  Ichii, Liu, Subramanian, Midgley, Miloslavich, Moln{\'a}r, Obura, Pfaff,
  Polasky, Purvis, Razzaque, Reyers, Chowdhury, Shin, {Visseren-Hamakers},
  Willis, and Zayas}]{diazGlobalAssessmentReport2019}
D{\'i}az SM, Settele J, Brond{\'i}zio E, et~al. (2019) The Global Assessment
  Report on Biodiversity and Ecosystem Services: {{Summary}} for Policy Makers.
  {Intergovernmental Science-Policy Platform on Biodiversity and Ecosystem
  Services}

\bibitem[{Dumas et~al.(2018{\natexlab{a}})Dumas, La~Rosa, Mendling, and
  Reijers}]{dumasFundamentalsBusinessProcess2018}
Dumas M, La~Rosa M, Mendling J, et~al. (2018{\natexlab{a}}) Fundamentals of
  {{Business Process Management}}. {Springer}, {Berlin, Heidelberg},
  \doi{10.1007/978-3-662-56509-4}

\bibitem[{Dumas et~al.(2018{\natexlab{b}})Dumas, La~Rosa, Mendling, and
  Reijers}]{dumasProcessRedesign2018}
Dumas M, La~Rosa M, Mendling J, et~al. (2018{\natexlab{b}}) Process
  {{Redesign}}, {Springer}, {Berlin, Heidelberg}, pp 297--339.
  \doi{10.1007/978-3-662-56509-4_8}

\bibitem[{Elkington(1998)}]{elkingtonACCOUNTINGTRIPLEBOTTOM1998}
Elkington J (1998) {{ACCOUNTING FOR THE TRIPLE BOTTOM LINE}}. Measuring
  Business Excellence 2(3):18--22. \doi{10.1108/eb025539}

\bibitem[{Ellram(1993)}]{ellramFrameworkTotalCost1993}
Ellram LM (1993) A {{Framework}} for {{Total Cost}} of {{Ownership}}. The
  International Journal of Logistics Management 4(2):49--60.
  \doi{10.1108/09574099310804984}

\bibitem[{{European Commission}(2014)}]{europeancommissionDirective2014952014}
{European Commission} (2014) Directive 2014/95/{{EU}} of the {{European
  Parliament}} and of the {{Council}} of 22 {{October}} 2014 amending
  {{Directive}} 2013/34/{{EU}} as regards disclosure of non-financial and
  diversity information by certain large undertakings and groups.
  \urlprefix\url{https://eur-lex.europa.eu/eli/dir/2014/95}

\bibitem[{{European Commission}(2020)}]{europeancommissionRegulation202085}
{European Commission} (2020) Regulation {(EU)} 2020/852 of the european
  parliament and of the council of 18 june 2020 on the establishment of a
  framework to facilitate sustainable investment, and amending regulation
  {(EU)} 2019/2088.
  \urlprefix\url{https://eur-lex.europa.eu/eli/reg/2020/852/oj}

\bibitem[{{European Commission} et~al.(2018){European Commission}, {Joint
  Research Centre}, Schau, Castellani, Fazio, Diaconu, Sala, Zampori, and
  Secchi}]{europeancommissionSupportingInformationCharacterisation2018}
{European Commission}, {Joint Research Centre}, Schau E, et~al. (2018)
  Supporting Information to the Characterisation Factors of Recommended {{EF
  Life Cycle Impact Assessment}} Methods : New Methods and Differences with
  {{ILCD}}. {Publications Office}, \doi{10.2760/671368}

\bibitem[{Feola(2020)}]{feolaCapitalismSustainabilityTransitions2020}
Feola G (2020) Capitalism in sustainability transitions research: {{Time}} for
  a critical turn? Environmental Innovation and Societal Transitions
  35:241--250. \doi{10.1016/j.eist.2019.02.005}

\bibitem[{Finkbeiner et~al.(2010)Finkbeiner, Schau, Lehmann, and
  Traverso}]{finkbeinerLifeCycleSustainability2010}
Finkbeiner M, Schau EM, Lehmann A, et~al. (2010) Towards {{Life Cycle
  Sustainability Assessment}}. Sustainability 2(10):3309--3322.
  \doi{10.3390/su2103309}

\bibitem[{Finnveden et~al.(2009)Finnveden, Hauschild, Ekvall, Guin{\'e}e,
  Heijungs, Hellweg, Koehler, Pennington, and
  Suh}]{finnvedenRecentDevelopmentsLife2009}
Finnveden G, Hauschild MZ, Ekvall T, et~al. (2009) Recent developments in
  {{Life Cycle Assessment}}. Journal of Environmental Management 91(1):1--21.
  \doi{10.1016/j.jenvman.2009.06.018}

\bibitem[{Freitas and Pereira(2015)}]{freitasProcessSimulationSupport2015}
Freitas AP, Pereira JLM (2015) Process simulation support in {{BPM}} tools:
  {{The}} case of {{BPMN}}

\bibitem[{Fritsch et~al.(2022)Fritsch, {von Hammerstein}, Schreiber, Betz, and
  Oberweis}]{fritschPathwaysGreenerPastures2022}
Fritsch A, {von Hammerstein} J, Schreiber C, et~al. (2022) Pathways to
  {{Greener Pastures}}: {{Research Opportunities}} to {{Integrate Life Cycle
  Assessment}} and {{Sustainable Business Process Management Based}} on a
  {{Systematic Tertiary Literature Review}}. Sustainability 14(18):11164.
  \doi{10.3390/su141811164}

\bibitem[{Georgoulakos et~al.(2017)Georgoulakos, Vergidis, Tsakalidis, and
  Samaras}]{georgoulakosEvolutionaryMultiObjectiveOptimization2017}
Georgoulakos K, Vergidis K, Tsakalidis G, et~al. (2017) Evolutionary
  {{Multi-Objective Optimization}} of business process designs with
  pre-processing. In: 2017 {{IEEE Congress}} on {{Evolutionary Computation}}
  ({{CEC}}). {IEEE}, {Donostia, San Sebasti\'an, Spain}, pp 897--904,
  \doi{10.1109/CEC.2017.7969404}

\bibitem[{Ghahfarokhi et~al.(2021)Ghahfarokhi, Park, Berti, and {van der
  Aalst}}]{ghahfarokhiOCELStandardObjectCentric2021}
Ghahfarokhi AF, Park G, Berti A, et~al. (2021) {{OCEL}}: {{A Standard}} for
  {{Object-Centric Event Logs}}. In: Bellatreche L, Dumas M, Karras P, et~al.
  (eds) New {{Trends}} in {{Database}} and {{Information Systems}}, vol 1450.
  {Springer International Publishing}, {Cham}, p 169--175,
  \doi{10.1007/978-3-030-85082-1_16}

\bibitem[{Giddings et~al.(2002)Giddings, Hopwood, and
  O'Brien}]{giddingsEnvironmentEconomySociety2002}
Giddings B, Hopwood B, O'Brien G (2002) Environment, economy and society:
  Fitting them together into sustainable development. Sustainable Development
  10(4):187--196. \doi{10.1002/sd.199}

\bibitem[{Glasson and
  Therivel(2019)}]{glassonIntroductionEnvironmentalImpact2019}
Glasson J, Therivel R (2019) Introduction to {{Environmental Impact
  Assessment}}, fifth edition edn. {Routledge}, {New York},
  \doi{10.4324/9780429470738}

\bibitem[{Goebel et~al.(1998)Goebel, Marshall, and
  Locander}]{goebelActivityBasedCosting1998}
Goebel DJ, Marshall GW, Locander WB (1998) Activity-{{Based Costing}}.
  Industrial Marketing Management 27(6):497--510.
  \doi{10.1016/S0019-8501(98)00005-4}

\bibitem[{Goedkoop(2007)}]{goedkoopEcoindicator99Methodology2007}
Goedkoop M (2007) The {{Eco-indicator}} 99 {{Methodology}}. Journal of Life
  Cycle Assessment, Japan 3(1):32--38. \doi{10.3370/lca.3.32}

\bibitem[{G{\"o}pel(2016)}]{gopelGreatMindshift2016}
G{\"o}pel M (2016) The {{Great Mindshift}}, The {{Anthropocene}}:
  {{Politik}}\textemdash{{Economics}}\textemdash{{Society}}\textemdash{{Science}},
  vol~2. {Springer International Publishing}, {Cham},
  \doi{10.1007/978-3-319-43766-8}

\bibitem[{Hern{\'a}ndez~Gonz{\'a}lez et~al.(2019)Hern{\'a}ndez~Gonz{\'a}lez,
  Calero, P{\'e}rez~Parra, and
  Mancebo}]{hernandezgonzalezApproachingGreenBPM2019}
Hern{\'a}ndez~Gonz{\'a}lez A, Calero C, P{\'e}rez~Parra D, et~al. (2019)
  Approaching {{Green BPM}} characterisation. Journal of Software: Evolution
  and Process 31(2):2145. \doi{10.1002/smr.2145}

\bibitem[{Herring(2006)}]{herringEnergyEfficiencyCritical2006}
Herring H (2006) Energy efficiency\textemdash a critical view. Energy
  31(1):10--20. \doi{10.1016/j.energy.2004.04.055}

\bibitem[{Houy et~al.(2011)Houy, Reiter, Fettke, and
  Loos}]{houyGreenBPMSustainability2011}
Houy C, Reiter M, Fettke P, et~al. (2011) Towards {{Green BPM}} \textendash{}
  {{Sustainability}} and {{Resource Efficiency}} through {{Business Process
  Management}}. In: {zur Muehlen} M, Su J (eds) Business {{Process Management
  Workshops}}, vol~66. {Springer}, {Berlin, Heidelberg}, p 501--510,
  \doi{10.1007/978-3-642-20511-8_46}

\bibitem[{Houy et~al.(2012)Houy, Reiter, Fettke, Loos, {Hoesch-Klohe}, and
  Ghose}]{houyAdvancingBusinessProcess2012}
Houy C, Reiter M, Fettke P, et~al. (2012) Advancing {{Business Process
  Technology}} for {{Humanity}}: {{Opportunities}} and {{Challenges}} of
  {{Green BPM}} for {{Sustainable Business Activities}}. In: Green {{Business
  Process Management}}. {Springer}, {Berlin, Heidelberg}, p 75--92,
  \doi{10.1007/978-3-642-27488-6_5}

\bibitem[{Hovorka et~al.(2012)Hovorka, Labajo, and
  Auerbach}]{hovorkaInformationSystemsEnvironmental2012}
Hovorka DS, Labajo E, Auerbach N (2012) Information {{Systems}} in
  {{Environmental Sustainability}}: {{Of Cannibals}} and {{Forks}}. In: vom
  Brocke J, Seidel S, Recker J (eds) Green {{Business Process Management}}.
  {Springer}, {Berlin, Heidelberg}, p 59--72, \doi{10.1007/978-3-642-27488-6_4}

\bibitem[{Howcroft and Trauth(2005)}]{howcroftHandbookCriticalInformation2005}
Howcroft D, Trauth E (2005) Handbook of {{Critical Information Systems
  Research}}. {Edward Elgar Publishing}, \doi{10.4337/9781845426743}

\bibitem[{Hueting and Reijnders(2004)}]{huetingBroadSustainabilityContra2004}
Hueting R, Reijnders L (2004) Broad sustainability contra sustainability: The
  proper construction of sustainability indicators. Ecological Economics
  50(3-4):249--260. \doi{10.1016/j.ecolecon.2004.03.031}

\bibitem[{{IEEE}(2016)}]{IEEEStandardEXtensible}
{IEEE} (2016) {{IEEE Standard}} for {{eXtensible Event Stream}} ({{XES}}) for
  {{Achieving Interoperability}} in {{Event Logs}} and {{Event Streams}}.
  \doi{10.1109/IEEESTD.2016.7740858}

\bibitem[{Ihde et~al.(2022)Ihde, Pufahl, V{\"o}lker, Goel, and
  Weske}]{ihdeFrameworkModelingExecuting2022}
Ihde S, Pufahl L, V{\"o}lker M, et~al. (2022) A framework for modeling and
  executing task-{{Specific}} resource allocations in business processes.
  Computing 104(11):2405--2429. \doi{10.1007/s00607-022-01093-2}

\bibitem[{{IPCC}(2015)}]{intergovernmentalpanelonclimatechangeClimateChange20142015}
{IPCC} (2015) Climate {{Change}} 2014: {{Mitigation}} of {{Climate Change}}:
  {{Working Group III Contribution}} to the {{IPCC Fifth Assessment Report}}.
  {Cambridge University Press}, {Cambridge}, \doi{10.1017/CBO9781107415416}

\bibitem[{ISO(2006)}]{ISO1404020062006}
ISO (2006) {{ISO}} 14040: 2006 - {{Environmental Management}}\textemdash{{Life
  Cycle Assessment}}\textemdash{{Principles}} and {{Framework}}. ISO

\bibitem[{Johnston et~al.(2007)Johnston, Everard, Santillo, and
  Rob{\`e}rt}]{johnstonReclaimingDefinitionSustainability2007}
Johnston P, Everard M, Santillo D, et~al. (2007) Reclaiming the definition of
  sustainability. Environmental science and pollution research international
  14(1):60--6. \doi{10.1065/espr2007.01.375}

\bibitem[{Klinkm{\"u}ller et~al.(2021)Klinkm{\"u}ller, Seeliger, M{\"u}ller,
  Pufahl, and Weber}]{klinkmulerDebugging2021}
Klinkm{\"u}ller C, Seeliger A, M{\"u}ller R, et~al. (2021) A method for
  debugging process discovery pipelines to analyze the consistency of model
  properties. In: Polyvyanyy A, Wynn MT, Van~Looy A, et~al. (eds) Business
  Process Management. Springer International Publishing, Cham, pp 65--84

\bibitem[{Kl{\"o}pffer et~al.(2014)Kl{\"o}pffer, Grahl, and
  Kl{\"o}pffer}]{klopfferLifeCycleAssessment2014}
Kl{\"o}pffer W, Grahl B, Kl{\"o}pffer W (2014) Life {{Cycle Assessment}}
  ({{LCA}}): A Guide to Best Practice. {Wiley-VCHVerlag GmbH \& Co. KGaA},
  {Weinheim}

\bibitem[{Kn{\"o}pfel et~al.(2005)Kn{\"o}pfel, Gr{\"o}ne, and
  Tabeling}]{knopfelFundamentalModelingConcepts2005}
Kn{\"o}pfel A, Gr{\"o}ne B, Tabeling P (2005) Fundamental Modeling Concepts:
  Effective Communication of {{IT}} Systems. {J. Wiley \& Sons}, {Chichester ;
  Hoboken, NJ}

\bibitem[{Lee and Kao(2001)}]{leeApplicationSimulationTechnique2001}
Lee TR, Kao JS (2001) Application of simulation technique to activity-based
  costing of agricultural systems: A case study. Agricultural Systems
  67(2):71--82. \doi{10.1016/s0308-521x(00)00042-1}

\bibitem[{Low et~al.(2016)Low, Vanden~Broucke, Wynn, Ter~Hofstede, De~Weerdt,
  and van Der~Aalst}]{lowRevisingHistoryCostinformed2016}
Low WZ, Vanden~Broucke SKLM, Wynn MT, et~al. (2016) Revising history for
  cost-informed process improvement. Computing 98(9):895--921.
  \doi{10.1007/s00607-015-0478-1}

\bibitem[{L{\"u}bbecke et~al.(2018{\natexlab{a}})L{\"u}bbecke, Fettke, and
  Loos}]{lubbeckeGuidelinesModelingEcologyAware2018}
L{\"u}bbecke P, Fettke P, Loos P (2018{\natexlab{a}}) Towards {{Guidelines}} of
  {{Modeling}} for {{Ecology-Aware Process Design}}. In: Teniente E, Weidlich M
  (eds) Business {{Process Management Workshops}}, vol 308. {Springer
  International Publishing}, {Cham}, p 510--519,
  \doi{10.1007/978-3-319-74030-0_40}

\bibitem[{L{\"u}bbecke et~al.(2018{\natexlab{b}})L{\"u}bbecke, Goswami, and
  Fettke}]{lubbeckeMethodEcologicalProcess2018}
L{\"u}bbecke P, Goswami A, Fettke P (2018{\natexlab{b}}) A {{Method}} for
  {{Ecological Process Optimization Based}} on {{Compliance Checking}}. In:
  2018 {{IEEE}} 20th {{Conference}} on {{Business Informatics}} ({{CBI}}).
  {IEEE}, {Vienna}, pp 119--128, \doi{10.1109/CBI.2018.00022}

\bibitem[{Mansar and Reijers(2005)}]{mansarBestPracticesBusiness2005}
Mansar SL, Reijers HA (2005) Best practices in business process redesign:
  Validation of a redesign framework. Computers in Industry 56(5):457--471.
  \doi{10.1016/j.compind.2005.01.001}

\bibitem[{Martin et~al.(2014)Martin, Depaire, and
  Caris}]{martinUseProcessMining2014}
Martin N, Depaire B, Caris A (2014) The use of process mining in a business
  process simulation context: {{Overview}} and challenges. In: 2014 {{IEEE
  Symposium}} on {{Computational Intelligence}} and {{Data Mining}} ({{CIDM}}).
  {IEEE}, {Orlando, FL, USA}, pp 381--388, \doi{10.1109/CIDM.2014.7008693}

\bibitem[{{Martinez-Alier} et~al.(1998){Martinez-Alier}, Munda, and
  O'Neill}]{martinez-alierWeakComparabilityValues1998}
{Martinez-Alier} J, Munda G, O'Neill J (1998) Weak comparability of values as a
  foundation for ecological economics. Ecological Economics 26(3):277--286.
  \doi{10.1016/S0921-8009(97)00120-1}

\bibitem[{Meredith et~al.(1989)Meredith, Raturi, Amoako-Gyampah, and
  Kaplan}]{meredithAlternativeResearchParadigms1989}
Meredith JR, Raturi A, Amoako-Gyampah K, et~al. (1989) Alternative research
  paradigms in operations. Journal of Operations Management 8(4):297--326.
  \doi{10.1016/0272-6963(89)90033-8}

\bibitem[{M{\"u}nstermann et~al.(2010)M{\"u}nstermann, Eckhardt, and
  Weitzel}]{munstermannPerformanceImpactBusiness2010}
M{\"u}nstermann B, Eckhardt A, Weitzel T (2010) The performance impact of
  business process standardization: {{An}} empirical evaluation of the
  recruitment process. Business Process Mgmt Journal 16(1):29--56.
  \doi{10.1108/14637151011017930}

\bibitem[{Naumann et~al.(2011)Naumann, Dick, Kern, and
  Johann}]{naumannGREENSOFTModelReference2011}
Naumann S, Dick M, Kern E, et~al. (2011) The {{GREENSOFT Model}}: {{A}}
  reference model for green and sustainable software and its engineering.
  Sustainable Computing: Informatics and Systems 1(4):294--304.
  \doi{10.1016/j.suscom.2011.06.004}

\bibitem[{Nowak et~al.(2011)Nowak, Leymann, Schumm, and
  Wetzstein}]{nowakArchitectureMethodologyFourPhased2011}
Nowak A, Leymann F, Schumm D, et~al. (2011) An {{Architecture}} and
  {{Methodology}} for a {{Four-Phased Approach}} to {{Green Business Process
  Reengineering}}. In: Kranzlm{\"u}ller D, Toja AM (eds) Information and
  {{Communication}} on {{Technology}} for the {{Fight}} against {{Global
  Warming}}, vol 6868. {Springer}, {Berlin, Heidelberg}, p 150--164,
  \doi{10.1007/978-3-642-23447-7_14}

\bibitem[{Nunamaker and
  Chen(1990)}]{nunamakerSystemsDevelopmentInformation1990}
Nunamaker J, Chen M (1990) Systems development in information systems research.
  In: Twenty-{{Third Annual Hawaii International Conference}} on {{System
  Sciences}}, vol iii. {IEEE Comput. Soc. Press}, {Kailua-Kona, HI, USA}, pp
  631--640, \doi{10.1109/HICSS.1990.205401}

\bibitem[{{OMG}(2011)}]{omg2011bpmn}
{OMG} (2011) Business process model and notation ({{BPMN}}), version 2.0

\bibitem[{Peffers et~al.(2007)Peffers, Tuunanen, Rothenberger, and
  Chatterjee}]{peffersDesignScienceResearch2007}
Peffers K, Tuunanen T, Rothenberger MA, et~al. (2007) A {{Design Science
  Research Methodology}} for {{Information Systems Research}}. Journal of
  Management Information Systems 24(3):45--77. \doi{10.2753/MIS0742-1222240302}

\bibitem[{Pufahl et~al.(2018)Pufahl, Wong, and
  Weske}]{pufahlDesignExtensibleBPMN2018}
Pufahl L, Wong TY, Weske M (2018) Design of an {{Extensible BPMN Process
  Simulator}}. In: Teniente E, Weidlich M (eds) Business {{Process Management
  Workshops}}, vol 308. {Springer International Publishing}, {Cham}, p
  782--795, \doi{10.1007/978-3-319-74030-0_62}

\bibitem[{Pufahl et~al.(2022)Pufahl, Zerbato, Weber, and
  Weber}]{pufahlBPMNHealthcareChallenges2022}
Pufahl L, Zerbato F, Weber B, et~al. (2022) {{BPMN}} in healthcare:
  {{Challenges}} and best practices. Information Systems 107:102013.
  \doi{10.1016/j.is.2022.102013}

\bibitem[{Purvis et~al.(2019)Purvis, Mao, and
  Robinson}]{purvisThreePillarsSustainability2019}
Purvis B, Mao Y, Robinson D (2019) Three pillars of sustainability: In search
  of conceptual origins. Sustainability Science 14(3):681--695.
  \doi{10.1007/s11625-018-0627-5}

\bibitem[{Rebmann et~al.(2022)Rebmann, Rehse, and van~der
  Aa}]{rebmanUncovering2024}
Rebmann A, Rehse JR, van~der Aa H (2022) Uncovering object-centric data
  in classical event logs for the automated transformation from xes
  to ocel. In: Di~Ciccio C, Dijkman R, del R{\'i}o~Ortega A, et~al. (eds)
  Business Process Management. Springer International Publishing, Cham, pp
  379--396

\bibitem[{Recker(2013)}]{reckerScientificResearchInformation2013}
Recker J (2013) Scientific {{Research}} in {{Information Systems}}: {{A
  Beginner}}'s {{Guide}}. {Springer}, {Berlin, Heidelberg},
  \doi{10.1007/978-3-642-30048-6}

\bibitem[{Recker et~al.(2011)Recker, Rosemann, and
  Gohar}]{reckerMeasuringCarbonFootprint2011}
Recker J, Rosemann M, Gohar ER (2011) Measuring the {{Carbon Footprint}} of
  {{Business Processes}}. In: Business {{Process Management Workshops}},
  vol~66. {Springer}, {Berlin, Heidelberg}, p 511--520,
  \doi{10.1007/978-3-642-20511-8_47}

\bibitem[{Recker et~al.(2012)Recker, Rosemann, Hjalmarsson, and
  Lind}]{reckerModelingAnalyzingCarbon2012}
Recker J, Rosemann M, Hjalmarsson A, et~al. (2012) Modeling and {{Analyzing}}
  the {{Carbon Footprint}} of {{Business Processes}}. In: vom Brocke J, Seidel
  S, Recker J (eds) Green {{Business Process Management}}. {Springer}, {Berlin,
  Heidelberg}, p 93--109, \doi{10.1007/978-3-642-27488-6_6}

\bibitem[{Romanelli et~al.(2015)Romanelli, Cooper, {Campbell-Lendrum}, Maiero,
  Karesh, Hunter, and Golden}]{romanelliConnectingGlobalPriorities2015}
Romanelli C, Cooper D, {Campbell-Lendrum} D, et~al. (2015) Connecting Global
  Priorities: Biodiversity and Human Health: A State of Knowledge Review.
  {World Health Organistion / Secretariat of the UN Convention on Biological
  Diversity}

\bibitem[{Roohy~Gohar and
  Indulska(2020)}]{roohygoharEnvironmentalSustainabilityGreen2020}
Roohy~Gohar S, Indulska M (2020) Environmental {{Sustainability}} through
  {{Green Business Process Management}}. Australasian Journal of Information
  Systems 24. \doi{10.3127/ajis.v24i0.2057}

\bibitem[{Satyal et~al.(2019)Satyal, Weber, Paik, {Di Ciccio}, and
  Mendling}]{satyalAB-BPM-Methodology2019}
Satyal S, Weber I, Paik H, et~al. (2019) Business process improvement with the
  {AB-BPM} methodology. Information Systems 84:283--298.
  \doi{10.1016/j.is.2018.06.007},
  \urlprefix\url{https://www.researchgate.net/publication/326256699_Business_Process_Improvement_with_the_AB-BPM_Methodology}

\bibitem[{Schneidewind and
  Zahrnt(2014)}]{schneidewindInstitutionalFrameworkSufficiency2014}
Schneidewind U, Zahrnt A (2014) The institutional framework for a sufficiency
  driven economy. \"Okologisches Wirtschaften - Fachzeitschrift 29(3):30.
  \doi{10.14512/OEW290330}

\bibitem[{Seidel et~al.(2012)Seidel, Recker, and vom
  Brocke}]{seidelGreenBusinessProcess2012}
Seidel S, Recker J, vom Brocke J (2012) Green {{Business Process Management}}.
  In: vom Brocke J, Seidel S, Recker J (eds) Green {{Business Process
  Management}}. {Springer}, {Berlin, Heidelberg}, p 3--13,
  \doi{10.1007/978-3-642-27488-6_1}

\bibitem[{Shove(2018)}]{shoveWhatWrongEnergy2018}
Shove E (2018) What is wrong with energy efficiency? Building Research \&
  Information 46(7):779--789. \doi{10.1080/09613218.2017.1361746}

\bibitem[{Sohns et~al.(2023)Sohns, Aysolmaz, Figge, and
  Joshi}]{sohnsGreenBusinessProcess2023}
Sohns TM, Aysolmaz B, Figge L, et~al. (2023) Green business process management
  for business sustainability: {{A}} case study of manufacturing small and
  medium-sized enterprises ({{SMEs}}) from {{Germany}}. Journal of Cleaner
  Production 401:136667. \doi{10.1016/j.jclepro.2023.136667}

\bibitem[{Stolze et~al.(2012)Stolze, Semmler, and
  Thomas}]{stolzeSustainabilityBusinessProcess2012}
Stolze C, Semmler G, Thomas O (2012) Sustainability in {{Business Process
  Management Research}}\textendash a {{Literature Review}}. AMCIS 2012
  Proceedings

\bibitem[{Supran et~al.(2023)Supran, Rahmstorf, and
  Oreskes}]{supranAssessingExxonMobilGlobal2023}
Supran G, Rahmstorf S, Oreskes N (2023) Assessing {{ExxonMobil}}'s global
  warming projections. Science 379(6628):eabk0063.
  \doi{10.1126/science.abk0063}

\bibitem[{Thies et~al.(2012)Thies, Dada, and
  {Stanoevska-Slabeva}}]{thiesPotentialNetworkCentricSolution2012}
Thies H, Dada A, {Stanoevska-Slabeva} K (2012) The {{Potential}} of a
  {{Network-Centric Solution}} for {{Sustainability}} in {{Business
  Processes}}. In: vom Brocke J, Seidel S, Recker J (eds) Green {{Business
  Process Management}}. {Springer}, {Berlin, Heidelberg}, p 181--201,
  \doi{10.1007/978-3-642-27488-6_11}

\bibitem[{{UN Environment}(2019)}]{unenvironmentGlobalEnvironmentOutlook2019}
{UN Environment} (ed)  (2019) Global {{Environment Outlook}} \textendash{}
  {{GEO-6}}: {{Summary}} for {{Policymakers}}. {Cambridge University Press},
  {Cambridge}, \doi{10.1017/9781108639217}

\bibitem[{{van der Aalst}(2016)}]{vanderaalstProcessMining2016}
{van der Aalst} WMP (2016) Process {{Mining}}. {Springer}, {Berlin,
  Heidelberg}, \doi{10.1007/978-3-662-49851-4}

\bibitem[{{van der Aalst}(2021)}]{vdaalstConcurrency2021}
{van der Aalst} WMP (2021) Concurrency and objects matter! disentangling the
  fabric of real operational processes to create digital twins. In: Cerone A,
  {\"O}lveczky PC (eds) Theoretical Aspects of Computing -- ICTAC 2021.
  Springer International Publishing, Cham, pp 3--17

\bibitem[{{van der Aalst} et~al.(2010){van der Aalst}, Nakatumba, Rozinat, and
  Russell}]{vanderaalstBusinessProcessSimulation2010}
{van der Aalst} WMP, Nakatumba J, Rozinat A, et~al. (2010) Business process
  simulation. In: vom Brocke J, Rosemann M (eds) Handbook on Business Process
  Management 1. International Handbooks on Information Systems, {Springer}, p
  313--338, \doi{10.1007/978-3-642-00416-2}

\bibitem[{{van der Kolk}(2022)}]{vanderkolkNumbersSpeakThemselves2022}
{van der Kolk} B (2022) Numbers {{Speak}} for {{Themselves}}, or {{Do They}}?
  {{On Performance Measurement}} and {{Its Implications}}. Business \& Society
  61(4):813--817. \doi{10.1177/00076503211068433}

\bibitem[{Vergidis and Tiwari(2008)}]{vergidisBusinessProcessDesign2008}
Vergidis K, Tiwari A (2008) Business process design and attribute optimization
  within an evolutionary framework. In: 2008 {{IEEE Congress}} on
  {{Evolutionary Computation}} ({{IEEE World Congress}} on {{Computational
  Intelligence}}). {IEEE}, {Hong Kong, China}, pp 668--675,
  \doi{10.1109/CEC.2008.4630867}

\bibitem[{Wernet et~al.(2016)Wernet, Bauer, Steubing, Reinhard, {Moreno-Ruiz},
  and Weidema}]{wernetEcoinventDatabaseVersion2016}
Wernet G, Bauer C, Steubing B, et~al. (2016) The ecoinvent database version 3
  (part {{I}}): Overview and methodology. The International Journal of Life
  Cycle Assessment 21(9):1218--1230. \doi{10.1007/s11367-016-1087-8}

\bibitem[{Weske(2012)}]{weskeBusinessProcessManagement2012}
Weske M (2012) Business {{Process Management}} - {{Concepts}}, {{Languages}},
  {{Architectures}}, 2nd edn. {Springer}, {Berlin, Heidelberg},
  \doi{10.1007/978-3-642-28616-2}

\bibitem[{West et~al.(2023)West, Wunder, Sills, B{\"o}rner, Rifai, Neidermeier,
  and Kontoleon}]{westActionNeededMake2023}
West TAP, Wunder S, Sills EO, et~al. (2023) Action needed to make carbon
  offsets from tropical forest conservation work for climate change mitigation.
  \doi{10.48550/ARXIV.2301.03354}

\bibitem[{Wesumperuma et~al.(2011)Wesumperuma, Ginige, Ginige, and
  Hol}]{wesumperumaFrameworkMultidimensionalBusiness2011}
Wesumperuma A, Ginige JA, Ginige A, et~al. (2011) A {{Framework}} for
  {{Multi-dimensional Business Process Optimization}} for {{GHG Emission
  Mitigation}}. ACIS 2011 Proceedings 91

\bibitem[{Wesumperuma et~al.(2013)Wesumperuma, Ginige, Ginige, and
  Hol}]{wesumperumaGreenActivityBased2013}
Wesumperuma A, Ginige A, Ginige JA, et~al. (2013) Green activity based
  management ({{ABM}}) for organisations. ACIS 2013 Proceedings

\bibitem[{Wieringa(2014)}]{wieringaDesignScienceMethodology2014}
Wieringa RJ (2014) Design {{Science Methodology}} for {{Information Systems}}
  and {{Software Engineering}}. {Springer}, {Berlin, Heidelberg},
  \doi{10.1007/978-3-662-43839-8}

\bibitem[{Wynn et~al.(2008)Wynn, Dumas, Fidge, {ter Hofstede}, and {van der
  Aalst}}]{wynnBusinessProcessSimulation2008}
Wynn MT, Dumas M, Fidge CJ, et~al. (2008) Business {{Process Simulation}} for
  {{Operational Decision Support}}. In: Hutchison D, Kanade T, Kittler J,
  et~al. (eds) Business {{Process Management Workshops}}, vol 4928. {Springer},
  {Berlin, Heidelberg}, p 66--77, \doi{10.1007/978-3-540-78238-4_8}

\bibitem[{Zeise et~al.(2012)Zeise, Link, and
  Ortner}]{zeiseMeasurementSystemsSustainability2012}
Zeise N, Link M, Ortner E (2012) Measurement {{Systems}} for
  {{Sustainability}}. In: {vom Brocke} J, Seidel S, Recker J (eds) Green
  {{Business Process Management}}. {Springer}, {Berlin, Heidelberg}, p
  131--146, \doi{10.1007/978-3-642-27488-6_8}

\bibitem[{Zeug et~al.(2023)Zeug, Bezama, and
  Thr{\"a}n}]{zeugLifeCycleSustainability2023}
Zeug W, Bezama A, Thr{\"a}n D (2023) Life {{Cycle Sustainability Assessment}}
  for {{Sustainable Bioeconomy}}, {{Societal-Ecological Transformation}} and
  {{Beyond}}. In: Hesser F, Kral I, Obersteiner G, et~al. (eds) Progress in
  {{Life Cycle Assessment}} 2021. Springer International Publishing, Cham, p
  131--159, \doi{10.1007/978-3-031-29294-1_8}

\end{thebibliography}

\end{document}